\documentclass{article}

\usepackage{amsmath,amssymb,amsfonts}
\usepackage{a4wide}
\usepackage{color}
\usepackage{xcolor}
\usepackage{bm}
\usepackage{graphicx}
\usepackage{euscript}
\usepackage{appendix}
\usepackage{lineno}

\newcommand{\cm}[1]{\color{black}{#1}}

\date{\today}
\title{
Boundary dissipative spin chains with partial solvability inherited from system Hamiltonians
}
\author{Chihiro Matsui
$^1$ and Naoto Tsuji$^{2,3}$ 
\\[3ex]
{\it $^1$Graduate School of Mathematical Sciences, The University of Tokyo}\\
{\it 3-8-1, Komaba, Meguro-ku, 153-8914 Tokyo, Japan}
\\
{\it $^2$Department of Physics, The University of Tokyo} \\
{\it 7-3-1, Hongo, Bunkyo-ku, 113-0033 Tokyo, Japan}\\
{\it $^3$RIKEN Center for Emergent Matter Science (CEMS)}\\
{\it 2-1, Hirosawa, Wako, 351-0198 Saitama, Japan}
}

\begin{document}
\maketitle

\begin{center}
{\bf Abstract}
\end{center}
\bigskip
{\small
Partial solvability plays an important role in the context of statistical mechanics, since it has turned out to be closely related to the emergence of quantum many-body scar states, i.e., exceptional energy eigenstates which do not obey 
the strong version of the eigenstate themalization hypothesis. 
We show that partial solvability of a quantum many-body system can be maintained even when the system is coupled to boundary dissipators under certain conditions. 
We propose two mechanisms that support partially solvable structures in boundary dissipative systems: The first one is based on the restricted spectrum generating algebra, while the second one is based on the Hilbert space fragmentation. 
From these structures,
we derive exact eigenmodes of the Gorini-Kossakowski-Sudarshan-Lindblad equation
for a family of quantum spin chain models with boundary dissipators,
where we find various intriguing phenomena arising from the partial solvability of the open quantum systems, including persistent oscillations (quantum synchronization) and the existence of the matrix product operator symmetry.
We discuss how the presence of solvable eigenmodes affects long-time behaviors of observables in boundary dissipative spin chains based on numerical simulations using the quantum trajectory method.
}

\section{Introduction}
Integrability of isolated quantum systems has been studied for a long time. Since the achievement by H. Bethe~\cite{bib:B31} who has derived the exact eigenfunctions for the Heisenberg spin chain, the method, now called the (coordinate) Bethe ansatz, has become a powerful tool for systematically constructing eigenfunctions of interacting many-body systems.
What lies behind the Bethe ansatz is the decomposability of a many-body scattering into ``consistent" two-body scatterings, that is, physics does not depend on a way to decompose a many-body scattering into two-body ones. This property, which is often referred to as the definition of ``quantum integrability", is guaranteed by the existence of the $R$-matrix that solves the Yang-Baxter equation (YBE)~\cite{bib:M64, bib:Y67}. 
Once the mathematical background of integrable systems has been understood, various methods have been invented for calculating energy spectra, form factors, correlation functions, etc. The methods include, e.g., the algebraic Bethe ansatz~\cite{bib:STF79, bib:FT84, bib:F96}, the vertex-operator approach~\cite{bib:DFJMN93, bib:JM95}, the separation of variables~\cite{bib:S85, bib:S90, bib:S92, bib:S92-2, bib:S95, bib:S96, bib:MN18}, and the off-diagonal Bethe ansatz~\cite{bib:CYSW13, bib:WYCS15, bib:ZCYSW14, bib:HCLYSW14}. 

On the other hand, there exists another class of solvable systems, in which only a part of the energy spectrum is analytically accessible. We shall call such a system ``partially solvable".
Partial solvability is mostly understood as an extra symmetry of the Hamiltonian that emerges only in a subspace $W$ of the entire Hilbert space $\mathcal H$. In other words, the Hamiltonian restricted in $W$ satisfies extra commutation relations induced by the extra symmetry. 
If the restricted Hamiltonian is integrable, 
such symmetry-induced commutation relations are derived from
infinitely many commuting transfer matrices originating from the YBE, leading to the existence of infinitely many conserved quantities. However, in most of the known cases of partially solvable systems, the extra symmetry is specified by a much 
simpler algebraic relation such as ``the restricted spectrum generating algebra (rSGA)"~\cite{bib:MRB20},
which holds in the subspace $W$. 
Examples of the partially solvable systems that exhibit the rSGA include the perturbed spin-$1$ $XY$ model~\cite{bib:SI19, bib:CIKM23} and the Affleck-Kennedy-Lieb-Tasaki (AKLT) model~\cite{bib:MRBR18, bib:MRBR18-2}. 
One can construct energy eigenstates in the solvable subspace
by applying the spectrum generating operator $Q$ (satisfying the rSGA with the Hamiltonian)  to a simply constructed energy eigenstate.

Another known mechanism which may induce partial solvability is the Hilbert space fragmentation (HSF)~\cite{bib:SRVKP20, bib:KHN20, bib:MPNRB21, bib:ZF21}. 
The HSF is defined as the fragmentation of the total Hilbert space into exponentially many invariant subspaces $\mathcal H_r$ (``Krylov subspaces") induced by the Hamiltonian,
\begin{equation} \label{eq:Krylov_subspace}
    \mathcal{H} = \bigoplus_r \mathcal{H}_r, 
\end{equation}
provided that such a decomposition is not caused by an obvious local symmetry of the Hamiltonian. 
\begin{figure} 
\begin{center}
    \includegraphics[width = 148mm]{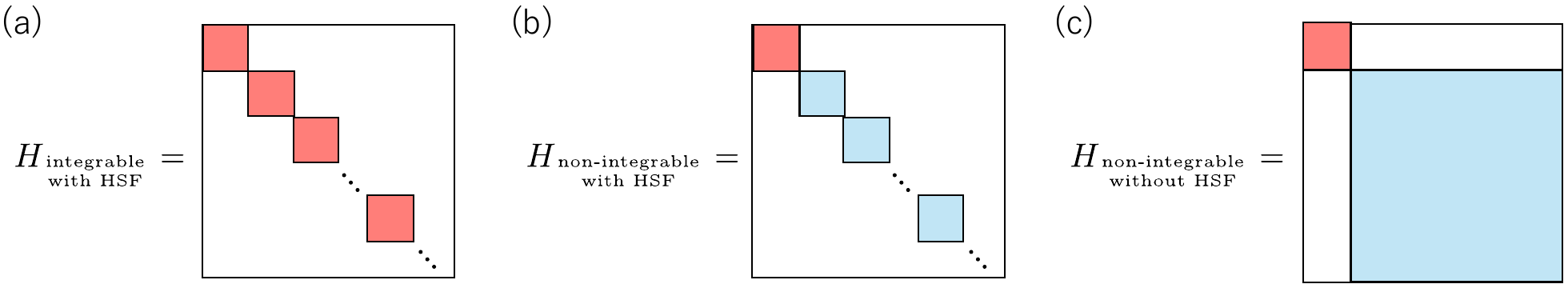}
    \caption{Fragmentation of the Hilbert space brought by three types of Hamiltonians. The Hamiltonian is integrable in the subspace colored with red, while non-interagrable in the subspaces colored with blue. (a) An integrable Hamiltonian with HSF. The Hilbert space consists of exponentially many subspaces forming (almost) block diagonal structure, in all of which the Hamiltonian is integrable. (b) A non-integrable Hamiltonian due to a perturbation which violates integrability but keeps the HSF structure. We choose a perturbation in such a way that keeps integrability in a selected subspace (colored with red). (c) A non-integrable Hamiltonian due to a perturbation which violates both integrability and the HSF structure. We show that it is still possible to choose a perturbation in such a way that keeps integrability in a selected subspace (colored with red). } \label{fig:HSF}
\end{center}
\end{figure}
The fragmentation structure is now mathematically understood in terms of ``the commutant algebra''\cite{bib:MM22} for the Hamiltonian, 
which also tells the number of Krylov subspaces induced by the Hamiltonian. 
Although the HSF is not necessary related to the notion of integrability, some models admit both the HSF and integrability (Fig.~\ref{fig:HSF}(a))~\cite{bib:Sato95, bib:M98, bib:M99}. A representative example is the $XXC$ model~\cite{bib:M98, bib:M99}, which has been introduced as a new type of integrable systems without referring to the HSF structure. 
As will be shown in \cite{bib:HMPP}, 
the Hamiltonian of the $XXC$ model has the fragmentation of the Hilbert space into exponentially many invariant subspaces, according to ``frozen" partial spin configurations (which we call ``irreducible strings (IS)" in the main text.)
Surprisingly, the integrable Hamiltonian with the HSF structure can be deformed in such a way that keeps its integrability in a selected subspace among exponentially many of those (Fig.~\ref{fig:HSF}(b), (c))~\cite{bib:HMPP}. The key idea is to choose a perturbation that vanishes
in a selected subspace and hence does not violate 
integrability in this subspace (Table~\ref{table:partially_intH}), although integrablity in the entire Hilbert space is in general lost by such a perturbation.
Based on a similar idea, we will introduce a variety of perturbations that keeps integrability in a selected subspace in the main text. A remarkable fact is that partial solvability in a selected subspace holds even under site-dependent integrability-breaking perturbations, since these perturbations are irrelevant in a selected subspace (a similar idea for Hamiltonians with rSGA can be found in \cite{bib:SYK20}). 
\begin{table} 
    \caption{Integrability in the entire Hilbert space $\mathcal{H}$ and its subspace $W$ for several systems. Adding an integrability-violating perturbation ($H_{{\rm pert}1}$) to an integrable Hamiltonian $H_{\rm int}$ generally breaks integrability in both $\mathcal{H}$ and $W$, while choosing a perturbation ($H_{\rm pert2}$) in such a way that $H_{{\rm pert}2}|_W = 0$ only breaks integrability in $\mathcal{H}$ but not in $W$.}  \label{table:partially_intH}
\begin{center}
    \begin{tabular}{r c c c}
    \hline\hline
    & \begin{tabular}{c} integrable \\ system \end{tabular} & \begin{tabular}{c} integrability-violating \\ perturbation \end{tabular} & \begin{tabular}{c} partially integrability- \\ preserving perturbation \end{tabular} \\
    \hline \vspace{-2mm} \\ 
    Hamiltonian & $H_{\rm int}$ & $H_{\rm int} + H_{{\rm pert}1}$ & $H_{\rm int} + H_{{\rm pert}2}$ \smallskip \\ 
    action restricted on $W$ & $H_{\rm int}|_W$ & $(H_{\rm int} + H_{{\rm pert}1})|_W \neq H_{\rm int}|_W$ & $H_{\rm int}|_W$ \vspace{2mm} \\
    integrability in $\mathcal{H}$ & $\checkmark$ & & \vspace{2mm} \\
    integrability in $W$ & $\checkmark$ & & $\checkmark$ \vspace{2mm} \\
    \hline\hline
    \end{tabular}
\end{center}
\end{table}

Partially solvable systems are now intensively studied especially in the context of thermalization. 
For instance, the solvable subspace $W$ is an invariant subspace of the Hamiltonian, and therefore, any state in $W$ never reaches the other subspaces during time evolution. This is a typical example of ``weak ergodicity breaking" in the Hilbert space, which may be considered as a necessary condition for the emergence of ``quantum many-body scar (QMBS) states"~\cite{bib:TMASP18}, i.e., non-thermal states in a non-integrable 
system. Indeed, many of the QMBS states have been found to be exactly solvable energy eigenstates of non-integrable systems. Several examples can be found in \cite{bib:MBR22}. 
Another remarkable feature in partially solvable systems is a persistent oscillation of local observables~\cite{bib:B2017, bib:TMASP18, bib:TMASP18-2, bib:AAW20, bib:HVRP20, bib:LMPC20, bib:HTC20, bib:CC21, bib:ZSMK21, bib:MTPAS20, bib:THSP19, bib:DHTP21}. This phenomenon is understood as a consequence of a large overlap between an initial state and solvable energy eigenstates forming equally-spaced energy spectra imposed by the rSGA.
Existence of long-lived oscillations implies that the system neither thermalizes nor relaxes
to any steady state. 

Motivated by these atypical behaviors of partially solvable systems, we focus on a question of whether an open quantum system can also admit partial solvability, and if so, what are characteristic phenomena in partially solvable open quantum systems. Let us consider open quantum systems that evolve according to the Gorini-Kossakowski-Sudarshan-Lindblad (GKSL) equation~\cite{bib:BP07}. 
The GKSL equation constitutes the most general form of the completely-positive-trace-preserving map under the assumptions that time evolution is Markovian (sometimes one also assumes an initial state to be a product state of the system and environment). 
Recent progress unveils several solvability mechanisms for the GKSL equation, which are classified into two types (Table \ref{tab:class_Liouvillians}): The first one consists of completely solvable Liouvillians whose spectra are fully accessible by analytic methods~\cite{bib:P08, bib:V20, bib:BBMJ20, bib:EP20, bib:MEP16, bib:SK19, bib:ZE20, bib:PP24}, while the second class consists of partially solvable Liouvillians whose spectra are only partially accessible~\cite{bib:P11, bib:MP17, bib:TMBJ20, bib:MT24, bib:WM24}. 
Most of the known partially solvable Liouvillians are solvable only for the steady state, leaving the other eigenmodes including the slowest decaying mode unsolvable (see, e.g., Ref.~\cite{bib:P11} and references therein). Only a few examples are known as partially solvable Liouvillians with a no-less-than-two-dimensional solvable subspace induced by rSGA \cite{bib:TMBJ20, bib:WM24}, but in those examples the dissipators are coupled to all the sites of the system.
\begin{table} \caption{
Several classes of solvable Liouvillians with bulk and boundary dissipators.
Completely solvable Liouvillians have been constructed both for bulk dissipators coupled to all the sites of the system~\cite{bib:P08, bib:V20, bib:BBMJ20, bib:EP20, bib:MEP16, bib:ZE20} and for boundary dissipators~\cite{bib:PP24}. Partial solvability of Liouvillians is mainly discussed only for the steady state~\cite{bib:P11}. A few examples which admit solvable eigenmodes have been found recently in \cite{bib:TMBJ20, bib:WM24}, although in those examples dissipators are attached at every site. In our work, we propose partially solvable Liouvillians with exactly solvable eigenmodes in the presence of boundary dissipators. } \label{tab:class_Liouvillians}
\begin{center}
    \begin{tabular}{r c c}
    \hline\hline
    & bulk dissipators & boundary dissipators \smallskip \\
    \hline \vspace{-2mm} \\ 
    completely solvable & $\checkmark$ & $\checkmark$ \vspace{2mm} \\
    steady state solvable & $\checkmark$ & $\checkmark$ \vspace{2mm} \\
    partially solvable & $\checkmark$ & our work \vspace{2mm} \\
    \hline\hline
    \end{tabular}
\end{center}
\end{table}

In this paper, we are especially interested in extending the notion of partial solvability for closed quantum systems to open quantum systems. Our target is an open quantum system with partial solvability inherited from a partially solvable system Hamiltonian, whose solvable energy eigenstates are robust against boundary dissipators.
One way to realize such an open quantum system is to employ a partially solvable system with rSGA, some of whose solvable eigenstates vanish by all the boundary dissipators.
We find that those solvable energy eigenstates may exist for the AKLT-type Hamiltonians when they couple to boundary dissipators injecting spin-2 quasiparticles. 
Another way is to employ a system Hamiltonian with the HSF structure and absorb the boundary dissipators as integrability-preserving
perturbations in a selected subspace. 
We find that there exist site-dependent perturbations that keep integrability in the selected subspace, and hence certain boundary dissipators can be interpreted as integrability-preserving
perturbations in the selected subspace of the integrable doubled spin chain (Fig.~\ref{fig:HSF}(c)). 
 As we will obtain in the main text, such a system may be realized for the perturbed $XXC$ model whose edges are attached to quasiparticle baths. 
In these partially solvable open quantum systems, the solvable subspace of the system Hamiltonian is 
inherited to a solvable subspace of the Liouvillian. Therefore, the partially solvable Liouvillians constructed in the above ways have not only the solvable steady states but also the solvable eigenmodes. As a result, we observe characteristic behaviors such as persistent oscillations, which can never be obtained for non-integrable Liouvillians. 

The rest of this paper is organized as follows. In Sec.~\ref{sec: closed system}, we give a review of partial solvability for closed quantum systems, and present our original results on site-dependent perturbations and an integrable subspace encoded by the period-three IS (see Sec.~\ref{sec:closed_HSF_pert}).
We explain two mechanisms of the partial solvability, the rSGA and HSF, together with some observations about the mechanism to incorporate the HSF and the integrability-preserving
perturbations in a selected subspace. Section~\ref{sec: open system} is the main part of this paper, which is devoted to partial solvability of open quantum systems. We show how the partial solvability of the system Hamiltonian can be inherited to open quantum systems, provided that the system shows either the rSGA or HSF. 
In the latter case, we take ``the thermofield double (TFD) formalism"~\cite{bib:I76, bib:M03}, which maps a density matrix defined in $\mathcal{H} \otimes \mathcal{H}^{\dag}$ to a vector in the doubled Hilbert space $\mathcal{H} \otimes \mathcal{H}^*$.
The action of the Liouvillian can be expressed as the Hamiltonian for the two decoupled integrable spin chains in the solvable subspace, if the quantum jump terms are irrelevant due to the HSF. 
We also demonstrate numerical results for some of the models to see how solvable eigenmodes in open quantum systems affect long-time behaviors of observables.
The concluding remarks are given in Sec.~\ref{sec: conclusion}, in which some open questions and possible future works are listed.

\section{Partially solvable closed spin chains}
\label{sec: closed system}

In this section, we mostly give an overview of partial solvability for closed quantum systems. 
In Sec.~\ref{sec:closed_HSF_pert}, we propose the existence of site-dependent perturbations to a certain integrable Hamiltonian that preserve integrability in a selected subspace, which is our original result. 
We mainly consider
$s=1$ spin chains with nearest-neighbor interactions as an example. The Hamiltonian can be written as
\begin{align} \label{eq:Hamiltonian}
    &H = \sum_{j=1}^{N} h_{j,j+1}, \qquad
    h_{j,j+1} = \bm{1} \otimes \cdots \otimes \underset{j,j+1}{h} \otimes \cdots \otimes \bm{1}, \\
	&h = \sum_{s,t,s',t'=0}^2 h^{s,s'}_{t,t'} |tt' \rangle \langle ss'|, \nonumber 
\end{align}
where $h$ is a local Hamiltonian acting on two neighboring $s=1$ spins whose states are labeled by $|tt'\rangle$ ($t,t'=0,1,2$), and $N$ is the number of lattice sites.
Among several mechanisms that produce partial solvability of closed quantum systems, we focus on the restricted spectrum generating algebra (rSGA) and the Hilbert space fragmentation (HSF). 

\subsection{Restricted spectrum generating algebra} \label{sec:closed_rSGA}

The notion of the spectrum generating algebra (SGA), or also called the dynamical symmetry, has been introduced in various contexts~\cite{bib:BB65, bib:DGN65, bib:BBN88, bib:SP98, bib:BO93, bib:L11}. In this paper, we shall say that the model has the SGA if there exists a spectrum generating operator $Q^{\dag}$ that satisfies an algebraic relation,  
\begin{align} \label{eq:SGA}
    [H,\,Q^{\dag}] - \mathcal{E} Q^{\dag} = 0,
\end{align}
for a real constant $\mathcal{E}$. 
If a model has the SGA and some of its energy eigenstates are known, one can construct towers of eigenstates by applying the operator $Q^{\dag}$ to those known states repeatedly. 
The observed energy spectrum is then equally spaced with the interval $\mathcal E$ due to the algebraic relation \eqref{eq:SGA}. 

One of the simplest examples that show the SGA is a free-fermion model. The Hamiltonian (denoted by $H_{\rm FF}$) is diagonalized in the momentum space,
\begin{align} \label{eq:free_fermion}
    H_{\rm FF} = \sum_k \Lambda_k \widetilde{\eta}_k^{\dag} \widetilde{\eta}_k,  
\end{align}
with real eigenvalues $\Lambda_k$ and a fermion creation operator $\widetilde{\eta}_k^{\dag}$,
and thus satisfies the SGA,
\begin{align}
    [H_{\rm FF},\,\widetilde{\eta}_k^{\dag}] = \Lambda_k \widetilde{\eta}_k^{\dag},
\end{align}
due to the anti-commutation relations for $\widetilde{\eta}_k^{\dag}$ and $\widetilde{\eta}_k$. 
Then all the energy eigenstates are created by applying the fermion creation operators $\widetilde{\eta}_k^{\dag}$ to the obvious vacuum state $|0 \rangle$, i.e., the zero-energy eigenstate,  
\begin{align}
    H_{\rm FF}\, \widetilde{\eta}_{k_1}^{\dag} \cdots \widetilde{\eta}_{k_n}^{\dag} |0 \rangle = (\Lambda_{k_1} + \cdots + \Lambda_{k_n})\, \widetilde{\eta}_{k_1}^{\dag} \cdots \widetilde{\eta}_{k_n}^{\dag} |0 \rangle.  
\end{align}
In this example, the fermion creation operator plays a role of the spectrum generating operator for each mode $k$. 

Sometimes there appears the SGA only in a subspace $W$ of the entire Hilbert space $\mathcal H$,
\begin{align} \label{eq:rSGA}
    [H,\,Q^{\dag}] - \mathcal{E} Q^{\dag} \big|_W = 0, \quad W \subset \mathcal{H}. 
\end{align}
We call this type of the SGA structure that emerges only in the subspace $W$ ``the restricted spectrum generating algebra (rSGA)". 
As in the case of the SGA that holds in the entire Hilbert space, one can construct a tower of energy eigenstates in the subspace $W$ by repeatedly applying the operator $Q^{\dag}$ to an obvious energy eigenstate $|\Psi_0 \rangle$ (if it may exist).

One of the simplest examples equipped with the rSGA is the perturbed spin-$1$ $XY$ model~\cite{bib:SI19}, whose local Hamiltonian is given by 
\begin{align}
    h_{j,j+1}^{XY} = \frac{J}{2} (S_j^+  S_{j+1}^- + S_j^-  S_{j+1}^+) + \frac{m}{2} (S_j^z  + S_{j+1}^z),
\end{align}
where $S_j^\pm=S_j^x\pm iS_j^y$ and $S_j^z$ are spin-1 operators, and $J$ and $m$ are real coefficients.
This model is considered to be non-integrable, and therefore, the energy eigenstates are in general not exactly solvable. However, one can easily find that the fully-polarized state $|22\dots 2 \rangle$ (and $|00\dots 0 \rangle$) is obviously an energy eigenstate with the eigenenergy $-mN$ ($mN$). Accordingly, one can derive some of the excited states exactly
in the form of $(Q^{\dag})^n |22\dots 2 \rangle$ (that allows a quasiparticle picture),
where $Q^\dagger$ creates quasiparticles of spin-2 magnons carrying momentum $k=\pi$
\begin{align} \label{eq:bimagnon}
    Q^{\dag} = \sum_{x=1}^N (-1)^{x} (S_x^+)^2. 
\end{align}
The spin-$2$ magnon creation operator $Q^{\dag}$ then satisfies the rSGA,
\begin{align} \label{eq:eSGA_XY}
    [H_{XY},\,Q^{\dag}] + 2m Q^{\dag} \big|_W = 0, 
\end{align}
in the subspace $W$
spanned by the quasiparticle excitation states
\begin{align}
    W = {\rm span}\{ (Q^{\dag})^n |22\dots 2 \rangle \}_{n \in \{0,1,\dots,N\}}.  
\end{align}
The rSGA produces the equally-spaced eigenvalue spectrum in the solvable subspace $W$, which can
explain the persistent oscillation observed in the Loschmidt echo~\cite{bib:SI19, bib:CIKM23}. This implies that the perturbed spin-$1$ $XY$ model never relaxes to any steady state, if the initial state has large enough overlap with the solvable subspace $W$. 

A more involved example exhibiting the rSGA is a family of the spin-$1$ Affleck-Kennedy-Lieb-Tasaki (AKLT)-type model~\cite{bib:AKLT87, bib:M24}. The local Hamiltonian is given by 
\begin{align} \label{eq:AKLT}
    h_{j,j+1}^{\rm AKLT} =& \frac{1}{2} h^{00}_{00} (S_j^x S_{j+1}^x + S_j^y S_{j+1}^y + S_j^z  S_{j+1}^z) \\
		&- \left( \frac{1}{2} h^{00}_{00} + \frac{a_1^2}{a_0 a_2} h^{11}_{11} \right) (S_j^x S_{j+1}^x + S_j^y S_{j+1}^y + S_j^z  S_{j+1}^z)^2 \nonumber  \\ 
		&- \left(  \frac{1}{2} h^{00}_{00} - \frac{a_1^2}{a_0 a_2} \Big(\frac{a_1^2}{a_0 a_2} - 1 \Big) h^{11}_{11} \right) (S_j^x S_{j+1}^x + S_j^y S_{j+1}^y)^2 \nonumber \\
		&- \left( h^{00}_{00} + \Big( \frac{a_1^2}{a_0 a_2} - 1 \Big) h^{11}_{11} \right) (S_j^z  S_{j+1}^z)^2 \nonumber \\
		&+ \left( \frac{1}{2} h^{00}_{00} + \Big( \frac{a_1^4}{a_0^2 a_2^2} - 1 \Big) h^{11}_{11} \right) ((S_j^z)^2 + (S_{j+1}^z)^2) 
		+ \left( 1 - 2 \frac{a_1^4}{a_0^2 a_2^2} \right) h^{11}_{11},  \nonumber 
\end{align}
in which $h^{00}_{00}$ and $h^{11}_{11}$ are free real parameters, while $a_0$, $a_1$, and $a_2$ are free complex parameters under the condition that $a_1^2/(a_0 a_2) \in \mathbb{R}$. Note that the original AKLT model is realized by choosing $h^{11}_{11} / h^{00}_{00} = 2/3$ and $a_0 = -\sqrt{2} a_1 = -a_2 = \sqrt{2/3}$. 
Although this Hamiltonian is non-integrable, it has been known for a long time that the ground state and some of the excited states are exactly solvable~\cite{bib:AKLT87, bib:A89}, to which a new list of solvable excited states has been added recently in the context of the QMBS states~\cite{bib:MBBFR20}. 

The zero-energy state
of the AKLT-type model is written in the form of the matrix product state,
\begin{align} \label{eq:MPS}
    &|\Psi_0 \rangle = \sum_{m_1,\dots m_N \in \{0,1,2\}} {\rm tr}_a (A_{m_1} \cdots A_{m_N}) |m_1 \dots m_N \rangle,
\end{align}
with the frustration-free condition
\begin{align}
    &h_{j,j+1}^{\rm AKLT} \vec{A}_j \vec{A}_{j+1} = 0, \quad
    \vec{A}_j = \begin{pmatrix} A_0 \\ A_1 \\ A_2 \end{pmatrix}_j,
\end{align}
which holds for $j = 1,2,\dots,N$. 
The matrix-valued elements $A_0$, $A_1$, and $A_2$ are given by the Pauli matrices 
\begin{align} \label{eq:MPS_elements}
    &A_0 = a_0 \sigma^+, \quad 
    A_1 = a_1 \sigma^Z, \quad
    A_2 = a_2 \sigma^-, \\
    &\sigma^{\pm} = \sigma^x \pm i \sigma^y, \nonumber
\end{align}
in which 
the coefficients $a_0$, $a_1$, and $a_2$ are the same as in the Hamiltonian \eqref{eq:AKLT}. Therefore, the bond dimension of this matrix product state is two. 

Besides the ground state, it has been found that several excited states are solvable, most of which admit the quasiparticle description~\cite{bib:CIKM23}. Here we focus on the excited states expressed by the spin-$2$ magnons carrying momentum $k = \pi$. They are created by the operator given in \eqref{eq:bimagnon}, 
which also satisfies the rSGA for the AKLT-type Hamiltonian,
\begin{align} \label{eq:rSGA_AKLT}
    &[H_{\rm AKLT},\,Q^{\dag}] - 2h^{00}_{00} Q^{\dag} \big|_W = 0,
\end{align}
in the subspace $W \subset \mathcal{H}$ spanned by the quasiparticle excited states 
\begin{align}
    &W = {\rm span}\{ (Q^{\dag})^n |\Psi_0 \rangle \}_{n \in \{0,1,\dots,\lfloor N/2 \rfloor\}}.  
\end{align}
Thus, the quasiparticle creation operator plays a role of the spectrum generating operator that creates a tower of solvable energy eigenstates on top of the matrix product 
zero-energy 
state \eqref{eq:MPS}. 

Recent studies on partial solvability have provided more formal understanding about the emergence of the rSGA in terms of quasisymmetry~\cite{bib:DBCK20, bib:DBCK20-2, bib:PPPK20, bib:PPPK21, bib:RLF21} for the former example and its deformation for the latter example~\cite{bib:RLF22}. 

%
\subsection{Hilbert space fragmentation}
Hilbert space fragmentation (HSF) is characterized by exponentially many block-diagonal structures of the Hamiltonian $H$, which are caused by non-obvious symmetries of $H$. Due to the block-diagonal structure of the Hilbert space, each subsapce is never accessed from the other subspaces by time evolution. 
Among several mechanisms for HSF~\cite{bib:PPN19, bib:SRVKP20, bib:KHN20}, we focus on the fragmentation observed for the Hilbert space of spin chains due to the presence of ``frozen" spin configurations under the action of a Hamiltonian. Those ``frozen" spin configurations are called  ``irreducible strings" (IS)~\cite{bib:DB93, bib:BD94, bib:MBD97}, which are associated with a non-obvious symmetry of the Hamiltonian.

In order to explain the HSF induced by IS, let us 
consider spin chains with arbitrary spin-$s$
(instead of spin-$1$).
Suppose that we have a spin-$s$ chain with nearest-neighbor interactions, whose local Hamiltonian is given by
\begin{align} \label{eq:local_h}
    h^{\rm HSF} &= \sum_{s,s' \in A} h^{s,s'}_{s,s'} |ss' \rangle \langle ss'| 
    + \sum_{t,t' \in B} h^{t,t'}_{t,t'} |tt' \rangle \langle tt'| \nonumber \\
    &\quad + \sum_{s \in A,\,t \in B} \Big(h^{s,t}_{t,s} |ts \rangle \langle st| + h^{t,s}_{s,t} |st \rangle \langle ts| + h^{s,t}_{s,t} |st \rangle \langle st| + h^{t,s}_{t,s} |ts \rangle \langle ts|\Big).
\end{align}
Here $A$ and $B$ represent subsets of the labels for local states $A,B \subset \{0,1,\dots 2s\}$, which satisfy $A \cup B = \{0,1,\dots 2s\}$ and $A \cap B = \emptyset$. We also call the labels for local states ``species". 
Since the Hamiltonian given in the form of Eq.~\eqref{eq:local_h} never exchanges the species in each subset $A$ or $B$, the configuration (i.e., IS) in each subset $A$ and $B$ is ``frozen" under the action of the Hamiltonian.
The existence of such a frozen partial configuration causes fragmentation of the Hilbert space, i.e., the the Hilbert space is fragmented according to the partial configurations in the subsets $A$ and $B$. 

\begin{figure} 
\begin{center}
    \includegraphics[width = 145mm]{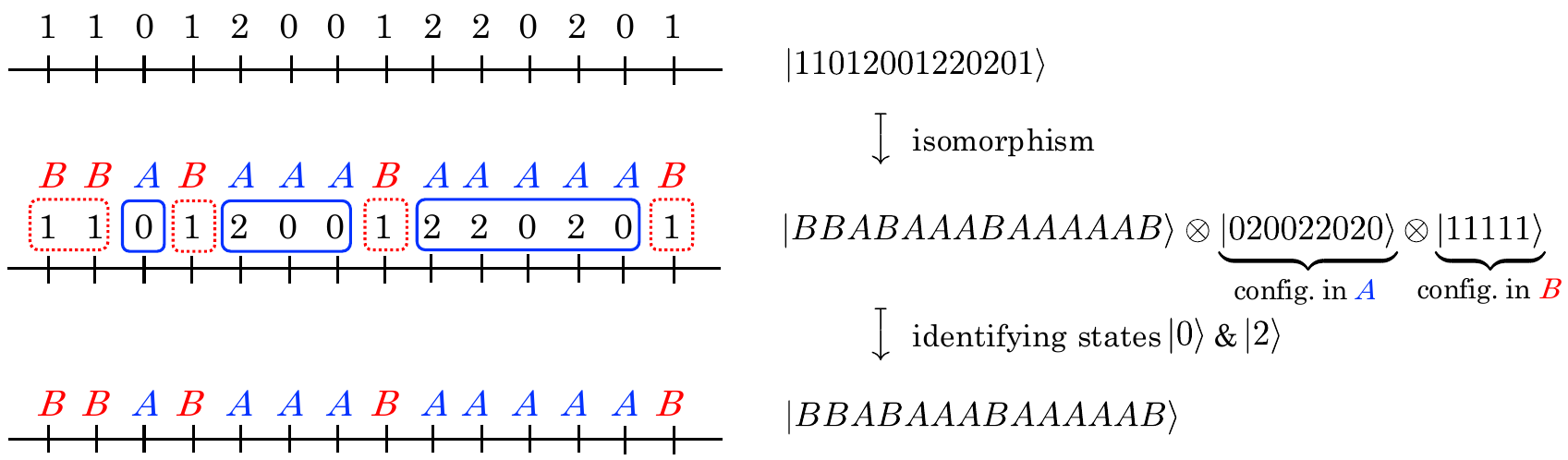}
    \caption{An example of IS formed by subsets $A = \{0,2\}$ and $B = \{1\}$. Given a state in the spin-1 representation (the first line), each local state belongs to either $A$ or $B$. The local states marked by the blue solid lines belong to $A$, while those marked with the red dotted lines belong to $B$. There exists an isomorphism which maps a state in the spin representation to a state expressed by two kinds of degrees of freedom (the second line), i.e., the labels $A$ and $B$, and configurations within $A$ and $B$ (which correspond to IS). For models with HSF induced by IS, the configurations in each subset are completely frozen, and therefore only the former degrees of freedom ($A$ and $B$) survive after identifying the local states $|0\rangle$ and $|2\rangle$ (the third line).} \label{fig:IS}
\end{center}
\end{figure}

In each of the fragmented subspaces, one can observe that a spin-$1/2$ model (with two local states) is embedded in the following way.
The entire Hilbert space of the spin-$s$ chain is $\mathbb{C}^{(2s+1)N}$, which consists of the tensor product of $N$ local linear spaces $\mathbb{C}^{2s+1}$ spanned by the $(2s+1)$ basis vectors $|0 \rangle, |1 \rangle,\dots,|2s+1 \rangle$ corresponding to the spin degrees of freedom. In this basis, a state in $\mathbb{C}^{(2s+1)N}$ is labeled by the spin configuration. 
Alternatively, one can use another basis 
with local states labeled by $A$ and $B$, and configurations realized within $A$ and $B$
(Fig.~\ref{fig:IS}). 
Suppose that the subset $A$ consists of $N_A$ species and $B$ consists of $N_B$ ($= (2s+1) - N_A$) species. The total degree of freedom is then calculated as 
\begin{align}
    \sum_{n=0}^N \begin{pmatrix} N \\ n \end{pmatrix} \cdot N_A^n N_B^{N-n} = (N_A +N_B)^N, 
\end{align}
which matches the dimension of the Hilbert space in the first description (${\rm dim}\,\mathbb{C}^{(2s+1)N}$), indicating that the two different bases describe the same Hilbert space. 
For the Hamiltonian with HSF \eqref{eq:local_h}, 
the configurations within $A$ and $B$ (i.e., IS) are frozen, while the labels $A$ and $B$ are unconstrained. The ($A, B$) degrees of freedom form the $N$-fold tensor product of the two-dimensional local linear space $\mathbb{C}^2$.
The projection onto such a $2^N$-dimensional subspace 
is defined by restricting to the configurations in $A$ and $B$ specified by IS (denoted by a projector $P_{\rm IS}$) and then by identifying all the local states in each of the subspaces $A$ and $B$ (Fig.~\ref{fig:IS}).
In this way, the Hamiltonian \eqref{eq:local_h} is reduced to a spin-$1/2$ chain with the nearest-neighbor interactions.


\subsubsection{Completely integrable case}
Although the HSF is not necessarily associated with the notion of solvability, it is often possible to embed solvability in one or a small number of subspaces. 
From now on, let us for simplicity come back to spin-$1$ systems.
One representative example is the spin-$1$ $XXC$ model~\cite{bib:M98, bib:M99} with the local Hamiltonian,
\begin{align} \label{eq:XXC}
    h^{XXC} = \cosh\eta \left(\sum_{s,s' \in \{0,2\}} |ss' \rangle \langle ss'| + |11 \rangle \langle 11| \right)
    + \sum_{s \in \{0,2\}} \left( |s1 \rangle \langle 1s| + |1s \rangle \langle s1| \right).  
\end{align}
The Hamiltonian exhibits the HSF, without changing the configuration of the labels $0$ and $2$, and thus the IS for the $XXC$ model is the partial configuration formed by $0$ and $2$. This local Hamiltonian indeed belongs to the class of \eqref{eq:local_h}, by choosing the subsets $A = \{0,2\}$ and $B = \{1\}$. 

Although the projector onto the subspaces specified by the IS generally has a site-dependent complicated expression, sometimes it can be expressed in a simple form.
One example is the projector onto the direct sum of the subspaces encoded by the fully-polarized ISs, $\dots 0000 \dots$ and $\dots 2222 \dots$, of length $n$ ($n=0,\dots,N$). The projectors onto the corresponding subspace are expressed by simple tensor products of the physical spaces,
\begin{align} \label{eq:proj_pol}
    &P^{(0)}_{\rm pol} = \bigotimes_{j=1}^N (|0 \rangle \langle 0| + |1 \rangle \langle 1|)_j, \qquad
    P^{(2)}_{\rm pol} = \bigotimes_{j=1}^N (|2 \rangle \langle 2| + |1 \rangle \langle 1|)_j,  
\end{align}
respectively.
Another example is the projector onto the subspace encoded by the alternating IS, $\dots 0202 \dots$~\cite{bib:HMPP},
\begin{align} \label{eq:proj_alt}
    &P_{\rm alt} 
    = \sum_{\genfrac{}{}{0pt}{}{(\alpha_1,\beta_1),\dots,(\alpha_N,\beta_N)}{\in \{0,1,2\}^2}}
    {\rm tr}_a (M_{\alpha_1,\beta_1} \cdots M_{\alpha_N,\beta_N})
    |\alpha_1 \dots \alpha_N \rangle \langle \beta_1 \dots \beta_N|
    - \bigotimes_{j=1}^N (|1 \rangle\langle 1|)_j, \\
    &M_{0,0} = \sigma_a^+, \quad M_{1,1} = \bm{1}_a, \quad M_{2,2} = \sigma_a^-, \quad M_{\alpha,\beta} = 0 \quad (\alpha \neq \beta), \nonumber
\end{align}
in which $\sigma_a^\pm$ and $\bm 1_a$ represent the Pauli matrices and the two-by-two unit matrix acting non-trivially on the auxiliary space, respectively. Here the trace ${\rm tr}_{a}$ is taken only over the auxiliary space.
This projector cannot be written in a simple tensor product form, but instead can be written in an (almost) matrix product form by introducing the two-dimensional auxiliary space. 
Note that the projectors $P_{\rm pol}^{(0)}$, $P_{\rm pol}^{(2)}$, and $P_{\rm alt}$ are not orthogonal to each other, but have a common one-dimensional subspace, ${\rm span}\{\otimes_{j=1}^N |1 \rangle_j\}$.

Besides the HSF structure, the $XXC$ model also exhibits integrability, which is guaranteed by the existence of the $R$-matrix solving the Yang-Baxter equation (YBE), 
\begin{align} \label{eq:YBE} 
    &R_{1,2}(\lambda_1,\lambda_2) R_{1,3}(\lambda_1,\lambda_3) R_{2,3}(\lambda_2,\lambda_3) = R_{2,3}(\lambda_2,\lambda_3) R_{1,3}(\lambda_1,\lambda_3) R_{1,2}(\lambda_1,\lambda_2)
    \quad (\lambda_1,\lambda_2,\lambda_3 \in \mathbb{C}), \nonumber    
\end{align}
defined in the three-fold tensor product of the linear spaces $V_1 \otimes V_2 \otimes V_3$. We denote the $R$-matrix that acts non-trivially on the $i$th and $j$th sites by $R_{i,j}$, e.g.,
\begin{align}
    R_{1,2}(\lambda_1,\lambda_2) = R(\lambda_1,\lambda_2) \otimes \bm{1}_3. 
\end{align}
The explicit form of the $R$-matrix for the $XXC$ model can be found in \cite{bib:M98}. 
Integrability of the $XXC$ model is inherited by the projected Hamiltonian onto the subspace specified by a certain IS. Regardless of the choice of IS, any projected Hamiltonian onto the subsapce specified by a certain IS results in the same spin-$1/2$ integrable $XXZ$ Hamiltonian by identifying $|0\rangle$ and $|2\rangle$, 
\begin{align}
    &h_{j,j+1}^{XXC} \underset{P_{\rm IS}\mathcal{H} \setminus \{ |0\rangle,|2\rangle \}^N}{\longmapsto} \cosh\eta \left( |\uparrow\uparrow \rangle \langle \uparrow\uparrow | + |\downarrow\downarrow \rangle \langle \downarrow\downarrow| \right)
    + \left( |\uparrow\downarrow \rangle \langle \downarrow\uparrow| + |\downarrow\uparrow \rangle \langle \uparrow\downarrow| \right),
\end{align}
where we assign up spins to the sites belonging to the subset $A$ and down spins to the sites belonging to the subset $B$. 
This also indicates that the spin-$1$ $XXC$ Hamiltonian can be diagonalized sector by sector labeled by an IS via the spin-$1/2$ $XXZ$ Hamiltonian, instead of being diagonalized directly via the $XXC$ $R$-matrix.

\subsubsection{Partially integrable case} \label{sec:closed_HSF_pert}

Now we consider perturbations that break entire integrability but keep integrability in a subspace specified by a given IS. 
%
The idea is to find integrability-breaking perturbations in such a way that are irrelevant (or vanish) in a given subspace. 

By focusing on systems with nearest-neighbor interactions, we provide two examples of perturbations that violate integrability in the entire Hilbert space but keep integrability in a given subspace. The first one is a perturbation which keeps integrability in the subspace specified by the polarized IS. 
In this subspace, any perturbation in a form of
\begin{align}
    h^{\rm pol}(\alpha_{{\rm d}1},\alpha_{{\rm d}2},\alpha_{{\rm o}1},\alpha_{{\rm o}2},\zeta) &= \alpha_{{\rm d}1} |02 \rangle \langle 02| + \alpha_{{\rm d}2} |20 \rangle \langle 20| + \alpha_{{\rm o}1} |02 \rangle \langle 20| + \alpha_{{\rm o}2} |20 \rangle \langle 02| 
    \label{eq:h^pol}
    \\
    &\quad + 2\cosh\zeta_1 \sum_{s \in \{0,2\}} |ss \rangle \langle ss|
    + 2\cosh\zeta_2 \sum_{s \in \{0,2\}} \left( |s1 \rangle \langle s1| + |1s \rangle \langle 1s| \right) \nonumber
\end{align}
does not violate integrability. Note that these perturbations also preserve the spin-flip invariance. It is easy to check that the first four interactions are irrelevant in the subspace of the polarized IS, since they vanish unless the configuration would include adjacent $0$ and $2$, which never appears in the polarized subspace. 
On the other hand, the interactions in the second line of Eq.~\eqref{eq:h^pol} act as a uniform external magnetic field in the projected space $P_{\rm pol} \mathcal{H} \setminus \{|0\rangle,|2\rangle\}^{N}$. 
Thus, the entire Hamiltonian acts in the projected space as the spin-$1/2$ $XXZ$ Hamiltonian with the shifted anisotropy $\cosh \eta \to \cosh\eta + \cosh\zeta_1 - \cosh\zeta_2$ and the uniform external magnetic field,
\begin{align} \label{eq:pert_pol}
    h_{j,j+1}^{XXC} + h_{j,j+1}^{\rm pol} \underset{P_{\rm pol}\mathcal{H} \setminus \{|0\rangle,|2\rangle\}^{N}}{\longmapsto}\,
    &\sigma_j^+ \sigma_{j+1}^- + \sigma_j^- \sigma_{j+1}^+ + \frac{1}{2}(\cosh\eta + \cosh\zeta_1 - \cosh\zeta_2)\, (\sigma_j^z \sigma_{j+1}^z) \\
    &+ \frac{1}{2}\cosh\zeta_1\, (\sigma_j^z  + \sigma_{j+1}^z) 
    + \frac{1}{2} (\cosh\eta\ + \cosh \zeta_1 + \cosh\zeta_2), \nonumber
\end{align}
which is integrable. 

The second example is a perturbation which keeps integrability in the subspace of alternating IS.
In this subspace, any of the following perturbations does not violate integrability,
\begin{align} \label{eq:pert_1}
    h^{\rm alt}(\beta_{{\rm d}1},\beta_{{\rm d}2},\beta_{{\rm o}1},\beta_{{\rm o}2},\zeta) &= \beta_{{\rm d}1} |00 \rangle \langle 00| + \beta_{{\rm d}2} |22 \rangle \langle 22| 
    + \beta_{{\rm o}1} |00 \rangle \langle 22| + \beta_{{\rm o}2} |22 \rangle \langle 00| \\
    &\quad + 2 \cosh\zeta_1 (|02 \rangle \langle 02| + |20 \rangle \langle 20|)
    + 2\cosh\zeta_2 \sum_{s \in \{0,2\}} \left( |s1 \rangle \langle s1| + |1s \rangle \langle 1s| \right). \nonumber
\end{align}
One can see that the first four terms are irrelevant, since they always vanish unless the state includes adjacent $0$s or $2$s, which never show up in the alternating subspace. On the other hand, the last two terms act as a uniform external magnetic field 
in the projected space $P_{\rm alt} \mathcal{H} \setminus \{|0\rangle,|2\rangle\}^{N}$. 
Thus, the entire Hamiltonian acts in this projected space as the spin-$1/2$ $XXZ$ model with the shifted anisotropy $\cosh \eta \to \cosh\eta + \cosh\zeta_1 - \cosh\zeta_2$ and the uniform external magnetic field,
\begin{align} \label{eq:pert_alt}
    h_{j,j+1}^{\rm XXC} + h_{j,j+1}^{\rm alt} \underset{P_{\rm alt}\mathcal{H} \setminus \{|0\rangle,|2\rangle\}^{N}}{\longmapsto}\,
    &\sigma_j^+ \sigma_{j+1}^- + \sigma_j^- \sigma_{j+1}^+ + \frac{1}{2}(\cosh\eta + \cosh\zeta_1 - \cosh\zeta_2)\, (\sigma_j^z  \sigma_{j+1}^z) \\
    &+ \frac{1}{2}\cosh\zeta\, (\sigma_j^z  + \sigma_{j+1}^z) 
    + \frac{1}{2} (\cosh\eta\ + \cosh \zeta_1 + \cosh\zeta_2),  \nonumber
\end{align}
which is again integrable. 
Note that the third and fourth terms in \eqref{eq:pert_1} violate not only integrability but also the spin-flip invariance, and consequently, the HSF structure of the entire Hilbert space. With these perturbations, total magnetization is no longer a conserved quantity. It is also worth notifying that both of the models \eqref{eq:pert_pol} and \eqref{eq:pert_alt} coincide with the $t$-$J_z$ model~\cite{bib:ZKMS97, bib:BO00, bib:LDOV23}, a canonical model that exhibits the HSF. 

Another remarkable fact is that partial solvability we discussed above is robust against site-dependent perturbations. For example, one can keep the system integrable in the subspace specified by the alternating IS even when one adds different perturbations on different sites, as long as the perturbations are written in the form of Eq.~\eqref{eq:pert_1}. We will come back to this point later in the discussion of partial solvability for open quantum systems. 

Although we have focused on nearest-neighbor interactions so far, 
there exist longer-range interactions that keep integrability in the subspace specified by an IS.
For instance, the following three-body interactions 
do not violate integrability in the subspace specified by the period-three triplet IS $(\dots 0\,0\,2\,0\,0\,2\,\dots)$:
\begin{equation} \label{eq:pert_2}
    h'_{\rm tri} = h^{000}_{000} |000 \rangle \langle 000| + h^{*22}_{*22} \bm{1} \otimes |22 \rangle \langle 22| + h^{22*}_{22*} |22 \rangle \langle 22| \otimes \bm{1} + h^{202}_{202} |202 \rangle \langle 202|. 
\end{equation}
In this way, integrability in the subspace specified by the IS with period-$p$ seems to hold under a certain choice of $p$-body interactions. Also, the projector onto the subspace encoded by the period-three triplet IS can be written in an (almost) matrix product form with a bond dimension three,
\begin{align}
    &P_{\rm tri} 
    = \sum_{\genfrac{}{}{0pt}{}{(\alpha_1,\beta_1),\dots,(\alpha_N,\beta_N)}{\in \{0,1,2\}^2}}
    {\rm tr}_a (M_{\alpha_1,\beta_1} \cdots M_{\alpha_N,\beta_N})
    |\alpha_1 \dots \alpha_N \rangle \langle \beta_1 \dots \beta_N|
    - 2 \bigotimes_{j=1}^N (|1 \rangle\langle 1|)_j, \\
    &M_{0,0} = S_a^+, \quad M_{1,1} = \bm{1}_a, \quad M_{2,2} = (S_a^-)^2, \quad M_{\alpha,\beta} = 0 \quad (\alpha \neq \beta), \nonumber
\end{align}
where $S_a^{\pm}$ and $\bm{1}_a$ represent the spin-$1$ operators and the three-by-three unit matrix acting non-trivially on the auxiliary space, respectively.

\subsubsection{Matrix product operator symmetry} \label{sec:MPO}

In the previous subsection, we have observed that the integrable spin-$1/2$ $XXZ$ model can be embedded in one of the fragmented Hilbert space of the $XXC$ model specified by a certain IS. Therefore, it is obviously possible to apply the Bethe ansatz method to construct partially conserved quantities, which are conserved only in the integrable subspace but not in the entire Hilbert space. On the other hand, it has been proposed that a partially solvable model is characterized by ``the matrix product operator (MPO) symmetry"~\cite{bib:BPP23}, which implies the existence of conserved quantities in the matrix product forms with fixed bond dimensions. Surprisingly, these are the conserved quantities not only in the integrable subspace but also in the entire Hilbert space. Since they have the matrix product forms with finite bond dimensions, they exhibit small entanglement entropies. 
Concrete examples have been displayed for the $XXC$ model~\cite{bib:BPP23} under the assumption that the conserved quantity is expressed by the MPO form,
\begin{align} \label{eq:MPO}
    &T = {\rm tr}_a (L_{a,N} \dots L_{a,1}), \\
    &L_{a,n} = \bm{1} \otimes |1 \rangle \langle 1| + \sum_{s,t=0,2} L^{(s,t)} \otimes |s \rangle \langle t|, \nonumber
\end{align}
in which $\bm{1}$ and $L^{(s,t)}$ are two-by-two matrices acting on the auxiliary space. 

The key relation in proving that the MPO \eqref{eq:MPO} commutes with the Hamiltonian is {\it the local divergence relation},
\begin{equation} \label{eq:local_divergence}
    [h_{j,j+1},\,L_{a,j+1} L_{a,j}] = M_{a,j+1} L_{a,j} - L_{a,j+1} M_{a,j},  
\end{equation}
where $M_{a,n}$ is another two-by-two matrix. 

For the $XXC$ model with the nearest-neighbor perturbations \eqref{eq:pert_1}, only $M = 0$ solves the divergence relation, which is included in the class of ``the commutant algebra" discussed in \cite{bib:MM22}. 
Two kinds of $L$ can be found as the solution to \eqref{eq:local_divergence}: The first one is the diagonal MPO,
\begin{align} \label{eq:MPO2}
    L^{(0,0)} = \begin{pmatrix} x & y \\ y & z \end{pmatrix}, \quad
    L^{(2,2)} = \begin{pmatrix} u & 0 \\ 0 & v \end{pmatrix}, 
\end{align}
where $x,y,z$ and $u,v$ are free parameters. The second one is the non-diagonal MPO, 
\begin{align}
    &L^{(0,2)} = (L^{(2,0)})^{\dag} = \gamma \sigma^-, \\
    &L^{(0,0)} = \begin{pmatrix} \alpha & 0 \\ 0 & \delta  \end{pmatrix}, \quad
    L^{(2,2)} = \begin{pmatrix} \beta & 0 \\ 0 & \varepsilon \end{pmatrix}, \nonumber
\end{align}
where $\alpha, \beta, \gamma, \delta, \varepsilon$ are free parameters. 

Note that the MPO symmetry with $M \neq 0$ has been found for the $XXC$ model with longer-range interactions \cite{bib:BPP23}, which is not included in the class of the commutant algebra~\cite{bib:MM22}.

\section{Partially solvable open spin chains}
\label{sec: open system}

Our main focus in this paper is partially solvable quantum systems coupled to boundary dissipators, in which the steady state and some eigenmodes are again solvable. In this section, we show two mechanisms to construct those models: 
The first one is an rSGA-induced partially solvable system which remains to be partially solvable even in the presence of boundary dissipators under a certain condition. The second one is the HSF-induced partially solvable system, whose HSF structure is partially inherited to the boundary dissipative system. 
These are thus examples of partially solvable boundary dissipative systems induced by partial solvability of the Hamiltonian, whose solvable states are robust against the boundary dissipators. 
In addition to robustness of partial solvability against boundary dissipators, there are several important questions including what are characteristic features of partially solvable eigemodes, how relaxation processes in the solvable subspace differ from those in a generic case, and how the partially solvable eigenmodes are experimentally realizable. 

In order to discuss these points, we consider a partially solvable spin-$1$ chain coupled to boundary dissipators, whose density matrix $\rho$ evolves according to the Liouvillian $\mathcal{L}$ in the GKSL equation,
\begin{align} \label{eq:Lindblad}
    &\frac{d}{dt} \rho(t) = \mathcal{L}(\rho) 
    = -i [H,\,\rho] + \sum_{\alpha} \gamma_{\alpha} \mathcal{D}_{\alpha}(\rho), \\
    &\mathcal{D}_{\alpha}(\rho) = A_{\alpha} \rho A_{\alpha}^{\dag} - \frac{1}{2} \{A_{\alpha}^{\dag} A_{\alpha},\,\rho\}. \nonumber
\end{align}
Here $H$ is the system's Hamiltonian and $A_{\alpha}$ is a quantum jump operator acting on the physical space. We assume that the jump operator $A_{\alpha}$ non-trivially acts only on the first and $N$th sites, representing the effect of a boundary dissipator with dissipation rates $\gamma_{\alpha}$. 
In the following discussion, we focus on two different Hamiltonians: the AKLT-type Hamiltonian \eqref{eq:AKLT} 
and the $XXC$ Hamiltonian \eqref{eq:XXC}.

%
\subsection{rSGA induced solvable eigenmodes}
The first example in which partial solvability is robust against boundary dissipators is given by an rSGA-induced partially solvable system. Let us consider a system in which a zero-energy state $|\Psi_0 \rangle$ is analytically known. The rSGA solvability is characterized by the existence of a spectrum generating operator $Q^{\dag}$ that satisfies the rSGA \eqref{eq:rSGA} with the Hamiltonian in a subspace $W$ spanned by states constructed by applying $Q^{\dag}$ to the zero-energy state,
\begin{align}
    W = \{ |\Psi_n \rangle \}_n, \quad
    |\Psi_n \rangle = (Q^{\dag})^n |\Psi_0 \rangle.
    \label{eq:Q^n}
\end{align}
Thus, the states $|\Psi_n \rangle$ are exactly solvable energy eigenstates of the Hamiltonian with the eigenenergies $\mathcal{E}n$. 
It often occurs that the solvable excited states admit the single-mode quasiparticle description,
\begin{align}
    Q^{\dag} = \sum_{x=1}^N e^{ikx} q^{\dag}_x, \nonumber 
\end{align}
in which $q_x^{\dag}$ is a local operator non-trivially acting on the $x$th site. 

Based on these, it is natural to ask whether the solvable states can survive even when quasiparticles are injected from both of the edges of the system.
Such a situation is realized by taking the boundary quantum jump operators as 
\begin{align}
    A_{\rm L} = q_1^{\dag}, \quad
    A_{\rm R} = q_N^{\dag}. 
\end{align}
The density matrix $\rho_{nn}$ for the solvable energy eigenstate $|\Psi_n \rangle$, as it commutes with the Hamiltonian by definition (i.e., $[\rho_{nn}, H]=0$), becomes the steady state of the GKSL equation \eqref{eq:Lindblad} if 
\begin{align} \label{eq:dark_state}
    &\mathcal{D}_{\alpha}(\rho_{nn}) = 0, \quad
    \forall \alpha, \\
    &\rho_{nn} = |\Psi_n \rangle \langle \Psi_n|. \nonumber
\end{align}
The pure state $\rho_{nn}$ that satisfies this condition together with the commutativity with the Hamiltonian is known as ``the dark state", which has been introduced in the context of atomic physics and optics~\cite{bib:DMKKBZ08, bib:KBDKMZ08}. 

The AKLT-type model \eqref{eq:AKLT} is one of the examples whose solvable energy eigenstates satisfy the dark-state condition \eqref{eq:dark_state} in the presence of the boundary quasiparticle dissipators.
The zero-energy eigenstate is given by the matrix product state, as has been explained in Sec.~\ref{sec:closed_rSGA}, but with the boundary deformation,
\begin{align} \label{eq:AKLT_open}
    |\Psi^{(v_{\rm L}, v_{\rm R})}_0 \rangle 
    = \sum_{m_1,\dots,m_N \in \{0,1,2\}}  {_a}\langle v_{\rm L}| A_{m_1} \cdots A_{m_N} |v_{\rm R} \rangle_a \cdot |m_1\dots m_N \rangle,
\end{align}
since the open boundary condition is imposed on the system. The boundary vectors $|v_{\rm L,R} \rangle \in V_a = {\rm span} \{ |\uparrow \rangle, |\downarrow \rangle \}$ in the auxiliary space must be properly chosen in order for \eqref{eq:AKLT_open} to be the zero-energy eigenstate. Especially when the model is frustration-free, as we consider in this paper, there are four degenerate zero-energy eigenstates, as no constraint is imposed on the boundary vectors. 
A tower of the solvable energy eigenstates are then independently constructed on top of each of the four degenerate zero-energy states by applying the spin-$2$ magnon creation operator $Q^{\dag}$ \eqref{eq:bimagnon}. That is, the solvable subspace $W$ under the open boundary condition is composed of four separate subspaces specified by the boundary vectors,
\begin{align} \label{eq:W_openrSGA}
    &W = W^{(\uparrow,\uparrow)} \oplus W^{(\uparrow,\downarrow)} \oplus W^{(\downarrow,\uparrow)} \oplus W^{(\downarrow,\downarrow)}, \\
    &W^{(v_{\rm L},v_{\rm R})} = {\rm span}\{ |\Psi_n^{(v_{\rm L},v_{\rm R})} \rangle \}_n, \quad 
    |\Psi_n^{(v_{\rm L},v_{\rm R})} \rangle = (Q^{\dag})^n |\Psi_0^{(v_{\rm L},v_{\rm R})} \rangle. \nonumber
\end{align}

We set the boundary quasiparticle dissipators in such a way that the quasiparticles are coming into the system from both of the ends,
\begin{align} \label{eq:diss_AKLT}
    A_{\rm L} = (S_1^+)^2, \quad 
    A_{\rm R} = (S_N^+)^2.     
\end{align}
With these dissipators, one of the solvable subspaces $W^{(\uparrow,\downarrow)}$ satisfies the dark state conditions \eqref{eq:dark_state},
\begin{align} \label{eq:dark_state2}
    &A_{\alpha} |\psi^{(\uparrow,\downarrow)} \rangle = 0, \quad \alpha \in \{{\rm L}, {\rm R}\}, \\
    &|\psi^{(\uparrow,\downarrow)} \rangle \in W^{(\uparrow,\downarrow)}. \nonumber
\end{align}
(See Appendix \ref{sec:proof} for the proof.) That is, by denoting the Liouvillian with the AKLT-type Hamiltonian and the dissipators \eqref{eq:diss_AKLT} by $\mathcal{L}_{\rm AKLT}$, any diagonal density matrix in the subspace $W^{(\uparrow,\downarrow)}$ 
becomes a steady state of the GKSL equation,
\begin{align}
    &\mathcal{L}_{\rm AKLT}(\rho_{\rm diag}^{(\uparrow,\downarrow)}) = 0, \\
    &\rho_{\rm diag}^{(\uparrow,\downarrow)} = \sum_n p_n |\Psi_n^{(\uparrow,\downarrow)} \rangle \langle \Psi_n^{(\uparrow,\downarrow)}|, \quad
    \sum_n p_n = 1, \, p_n \geq 0,\, \forall n. \nonumber
\end{align}
Note that, in the case of the perturbed spin-$1$ $XY$ model, which is another model with the hidden rSGA, there never exist such a solvable energy eigenstate that satisfies the dark-state condition \eqref{eq:dark_state2}.  

The fact that any state in the subspace $W^{(\uparrow,\downarrow)}$ is the dark state \eqref{eq:dark_state2} indicates that any density matrix in $W^{(\uparrow,\downarrow)} \otimes (W^{(\uparrow,\downarrow)})^{\dag}$, even if it contains off-diagonal elements, becomes an eigenmode of the GKSL equation. Suppose that we have an off-diagonal element $\rho_{m,n}^{(\uparrow,\downarrow)} = |\Psi_m^{(\uparrow,\downarrow)} \rangle \langle \Psi_n^{(\uparrow,\downarrow)}|$ in a given density matrix. This is a dark state from the previous statement, and moreover, its commutator with the Hamiltonian is proportional to itself,
\begin{align} \label{eq:SGA_GKSL}
    [H,\, \rho_{m,n}^{(\uparrow,\downarrow)}] = 2(m-n) h^{00}_{00} \rho_{m,n}^{(\uparrow,\downarrow)},
\end{align}
which indicates that $\rho_{m,n}^{(\uparrow,\downarrow)}$ is an eigenmode of the GKSL equation. The relation \eqref{eq:SGA_GKSL} together with the dark-state condition \eqref{eq:dark_state2} leads to the restricted spectrum generating algebra for the Liouvillian in the subspace $W^{(\uparrow,\downarrow)} \otimes (W^{(\uparrow,\downarrow)})^{\dag} \subset \mathcal{H} \otimes \mathcal{H}^{\dag}$,
\begin{align} \label{eq:SGA_GKSL2}
    \mathcal{L}_{\rm AKLT}(\rho^{(\uparrow,\downarrow)}_{m,n})
    = -2i (m-n) h^{00}_{00} \rho^{(\uparrow,\downarrow)}_{m,n}, 
\end{align}
giving an equally-spaced spectrum along the imaginary axis with the interval $2h^{00}_{00}$ embedded in the full spectrum of $\mathcal{L}_{\rm AKLT}$. 

The hidden rSGA structure \eqref{eq:SGA_GKSL2} of the Liouvillian evokes us the persistent oscillations observed for rSGA-induced partially solvable isolated quantum systems~\cite{bib:TMASP18}. Indeed, if we choose an initial state in the subspace $W^{(\uparrow,\downarrow)}$,
\begin{align}
    |\psi(0) \rangle = \sum_n a_n |\Psi_n^{(\uparrow,\downarrow)} \rangle \in W^{(\uparrow,\downarrow)}, 
\end{align}
it is easy to show that persistent oscillations are observed for an observable $O$ in a long-time scale,
\begin{align}
    \langle O(t) \rangle
    \sim \sum_{n \leq m} 2\cos(2(m-n)h^{00}_{00} t) a_m a_n\, {\rm Re}\, O_{nm} \quad (t\to\infty). 
\end{align}
This indicates that the system prepared in the solvable subspace never relaxes to any steady state. 
Even if the initial state is generic, the long-lived oscillations survive as long as the initial state has a large enough overlap with the subspace $W^{(\uparrow,\downarrow)}$. 

\subsection{Numerical simulation}

To see how the presence of the rSGA-induced solvable eigenmodes affects observables in partially solvable open spin chains,
here we perform numerical simulations for the GKSL equation (\ref{eq:Lindblad}).
We take the generalized AKLT model $H_{\rm AKLT}=\sum_{j=1}^{N-1} h_{j,j+1}^{\rm AKLT}$ (\ref{eq:AKLT}) as the bulk Hamiltonian
with a finite number of lattice sites $N$. We specifically focus on the case of the original AKLT model, i.e., by choosing
$h_{11}^{11}/h_{00}^{00}=2/3, a_0=-\sqrt{2}a_1=-a_2=\sqrt{2/3}$.
The dissipators are given by spin-2 creation operators (\ref{eq:diss_AKLT})
acting on each end of the chain. The explicit form of the GKSL equation that we solve in this section is:
\begin{align}
\frac{d}{dt}\rho
&=
-i[H_{\rm AKLT},\rho]+\sum_{\alpha={\rm L,R}} \gamma_\alpha \left( A_\alpha \rho A_\alpha^\dagger - \frac{1}{2}\{A_\alpha^\dagger A_\alpha, \rho\}\right),
\\
H_{\rm AKLT}
&=
\sum_{j=1}^{N-1}
\left[
\bm S_j \cdot \bm S_{j+1}+\frac{1}{3} (\bm S_j \cdot \bm S_{j+1})^2
\right],
\\
A_{\rm L}
&=
(S_1^+)^2,
\quad
A_{\rm R}
=
(S_N^+)^2,
\end{align}
which is numerically simulated by the quantum trajectory method \cite{bib:DCM92,bib:Carmichael93,bib:Daley14} together with the exact diagonalization.

The initial state is chosen to be a product state of the spin-1 chain, whose spin configuration is nearly a N\'eel state $|N, S^z(0)\rangle$,
depending on the values of $N$ and the initial total $S^z(0)$:
\begin{align}
|0 2 0 2 \cdots 0 2 0 2 \rangle : 
&\quad
\mbox{$N$ is even,\, $S^z(0)=0$}
\label{eq:Neel1}
\\
|0 2 0 2 \cdots 0 2 0 1 \rangle : 
&\quad
\mbox{$N$ is even,\, $S^z(0)=1$}
\label{eq:Neel2}
\\
|0 2 0 2 \cdots 0 2 0 2 1 \rangle : 
&\quad
\mbox{$N$ is odd,\, $S^z(0)=0$}
\label{eq:Neel3}
\\
|0 2 0 2 \cdots 0 2 0 2 0 \rangle : 
&\quad
\mbox{$N$ is odd,\, $S^z(0)=1$}
\label{eq:Neel4}
\end{align}
We identify the state labels $0, 1, 2$ with the spin configuration $\uparrow, 0, \downarrow$, respectively. Among the initial states $|N, S^z(0)\rangle$ considered here, those with $S^z(0)=1$ have an overlap with the states in the subspace $W^{(\uparrow,\downarrow)}$, since they have $\uparrow$ spins at both ends of the chain after removing $S_j^z=0$ spins. Hence, we expect that the rSGA-induced solvable eigenmodes appear in those cases with $S^z(0)=1$.

\begin{figure}[t]
\centering
\includegraphics[width=7cm]{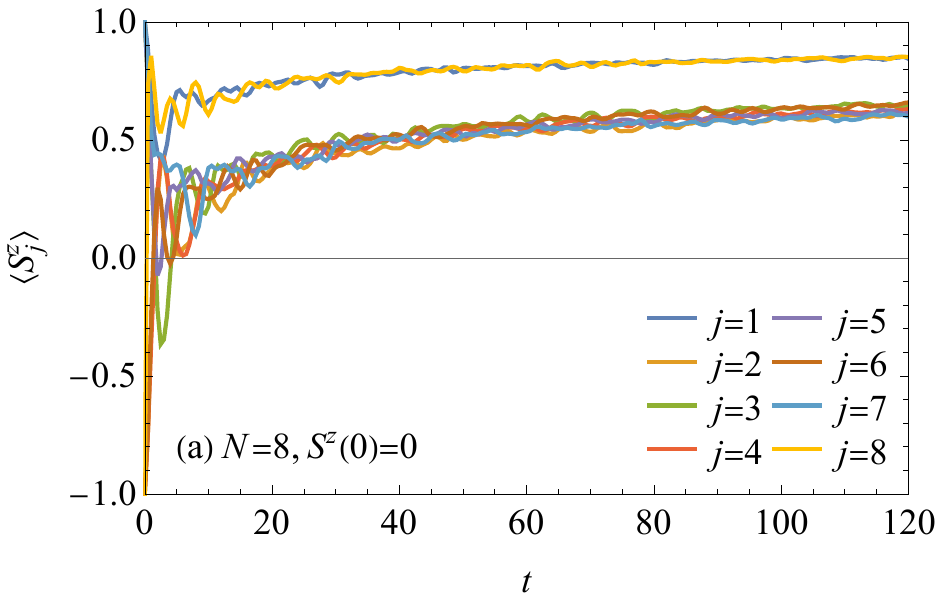}
\includegraphics[width=7cm]{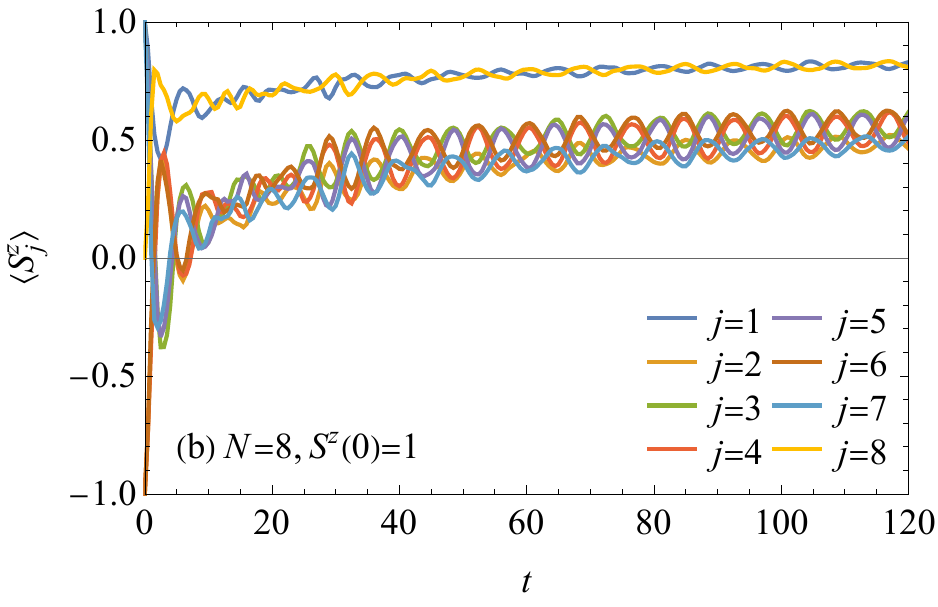}
\includegraphics[width=7cm]{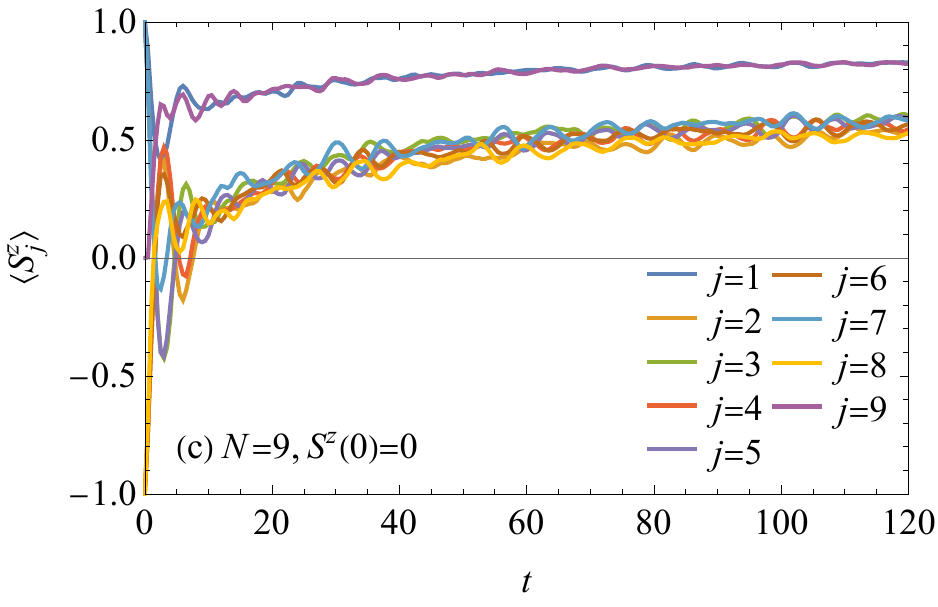}
\includegraphics[width=7cm]{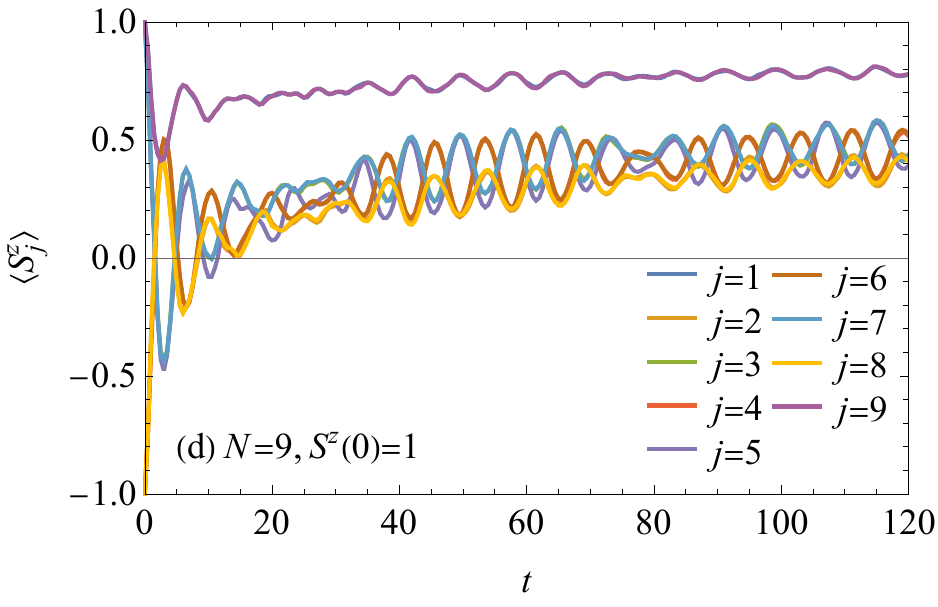}
\caption{Time evolution of the local magnetization $\langle S_j^z\rangle$ for the dissipative AKLT model
with (a) $N=8, S^z(0)=0$, (b) $N=8, S^z(0)=1$, (c) $N=9, S^z(0)=0$, and (d) $N=9, S^z(0)=1$.
The initial state is a nearly N\'eel state defined by Eqs.~(\ref{eq:Neel1})-(\ref{eq:Neel4}).
The dissipation rate is taken to be $\gamma_{\rm L}=\gamma_{\rm R}=1$.
We take an average over 1000 trajectories.}
\label{fig:Sz}
\end{figure}

In Fig.~\ref{fig:Sz}, we show the time evolution of the local magnetization $\langle S_j^z\rangle$ for several initial conditions
((a) $N=8, S^z=0$, (b) $N=8, S^z=1$, (c) $N=9, S^z=0$, (d) $N=9, S^z=1$)
with $\gamma_{\rm L}=\gamma_{\rm R}=1$. In the cases of $S^z(0)=0$ [Fig.~\ref{fig:Sz}(a), (c)], the local magnetization gradually reaches 
the steady state value without oscillations, while in the cases of $S^z(0)=1$ [Fig.~\ref{fig:Sz}(b), (d)]
the local magnetization clearly shows long-lived coherent oscillations. In all the cases above, the magnetization does not approach the maximum value, meaning that the steady state does not correspond to the trivial all up states (i.e., $|00\cdots 0\rangle=|\uparrow\uparrow \cdots \uparrow\rangle$). We can also see that the magnetization at the boundary $j=1, N$ takes a relatively larger steady-state value as compared to the bulk part, which may be due to the effect of the spin injection at the boundary.

\begin{figure}[t]
\centering
\includegraphics[width=7cm]{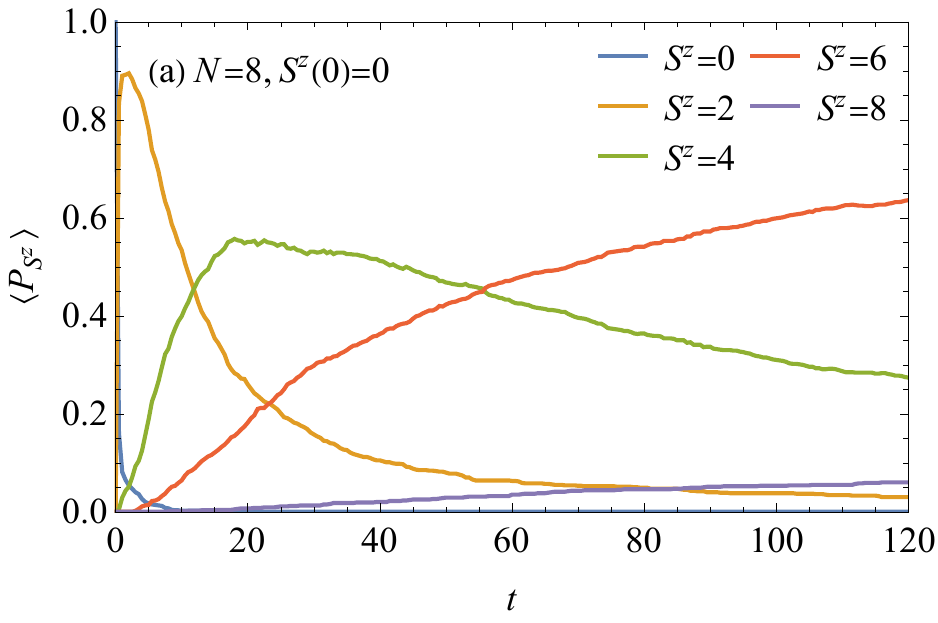}
\includegraphics[width=7cm]{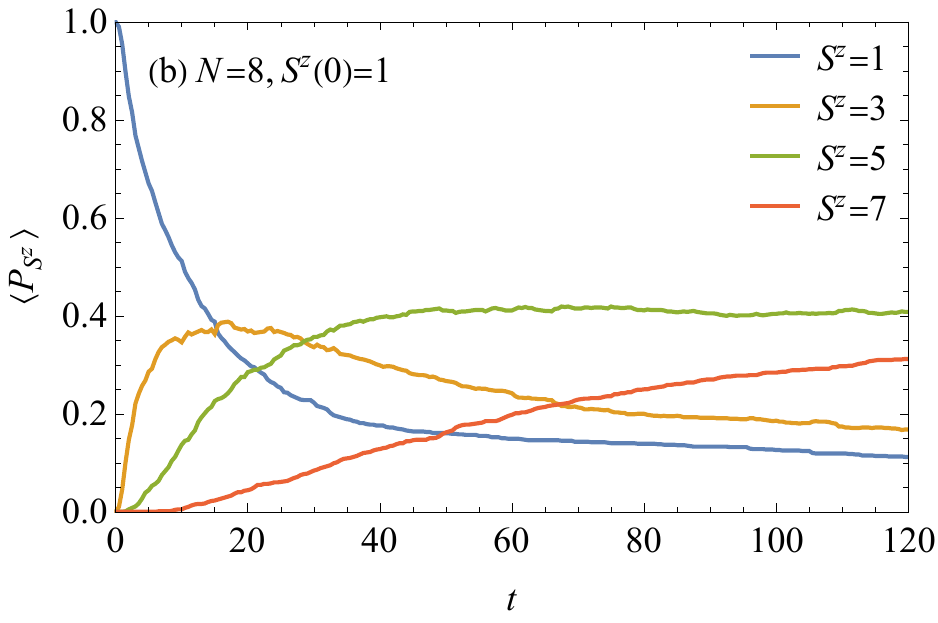}
\includegraphics[width=7cm]{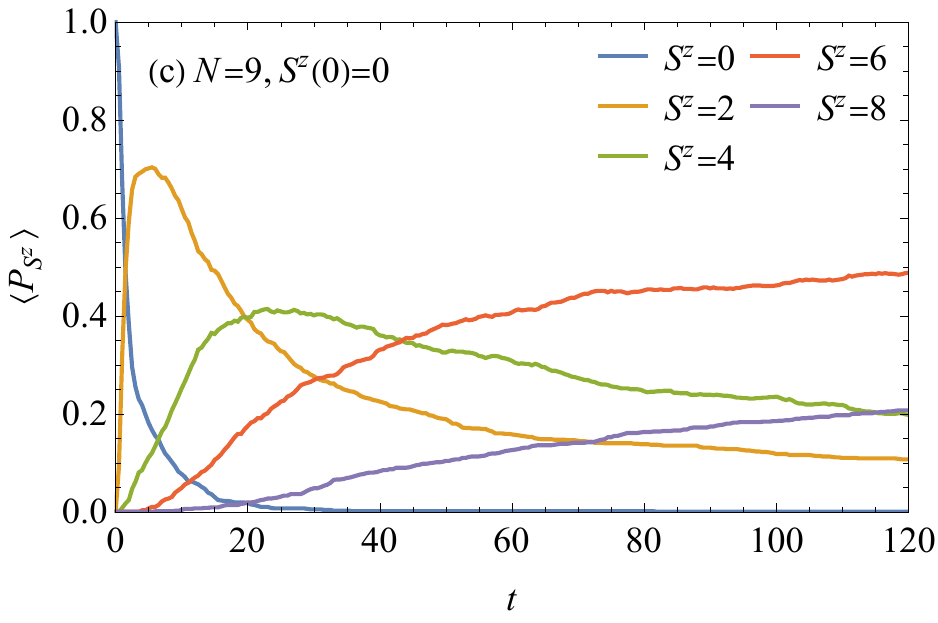}
\includegraphics[width=7cm]{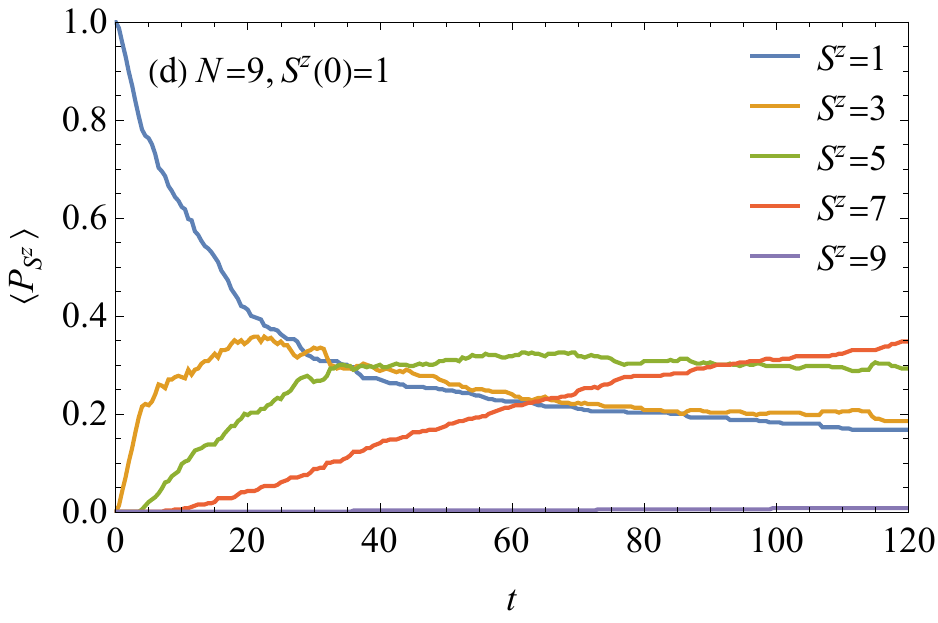}
\caption{The ratio of the number of trajectories for each $S^z$ in the dissipative AKLT model as a function of time
with (a) $N=8$ and the initial $S^z=0$, (b) $N=8$ and the initial $S^z=1$, (c) $N=9$ and the initial $S^z=0$, and (d) $N=9$ and the initial $S^z=1$.
The initial state is a nearly N\'eel state defined by Eqs.~(\ref{eq:Neel1})-(\ref{eq:Neel4}).
The dissipation rate is taken to be $\gamma_{\rm L}=\gamma_{\rm R}=1$.
We take an average over 1000 trajectories.}
\label{fig:P}
\end{figure}

In order to understand the role of the solvable eigenmodes in the GKSL equation, we plot the ratio of the number of trajectories for each $S^z$ measured by $\langle P_{S^z}\rangle$ in Fig.~\ref{fig:P}, where $P_{S^z}$ is the projection operator onto the corresponding subspace with fixed $S^z$. The parameters are the same as in Fig.~\ref{fig:Sz}. When $S^z(0)=0$ [Fig.~\ref{fig:P}(a), (c)], the number of trajectories having $S^z=0$ quickly decays to zero, while those with $S^z\neq 0$ grow subsequently (those with lower $S^z$ grow faster). Since $S^z$ can change by 2 due to the spin-2 injection at the boundaries, $S^z$ only takes values of even integers (that should not exceed the system size $N$). In the long-time limit, 
the trajectories with $S^z=6$ and $8$ survive, and the others seem to vanish.
This is in sharp contrast to the cases for $S^z(0)=1$ [Fig.~\ref{fig:P}(b), (d)],
where the number of trajectories with arbitrary $S^z$ can survive in the long-time limit. This is in consistent with the fact that there is a tower of dark states with $S^z=1,3,5,\dots$ (as shown in Eq.~(\ref{eq:Q^n})), which can be accessed from the initial state with $S^z(0)=1$ having an overlap with the states in the subspace $W^{(\uparrow,\downarrow)}$. Hence the steady state remains to be far from the trivial all up states ($|00\cdots 0\rangle=|\uparrow\uparrow \cdots \uparrow\rangle$) even in the presence of the spin-2 injection. The steady states for $S^z(0)=1$ are also different from the all up state, since there is an additional dark state with $S^z=S_{\rm max}^z-2$ ($S_{\rm max}^z$ is the maximum $S^z$ in a finite chain with length $N$). In the thermodynamic limit, this dark state will become indistinguishable from the all up state.
In this way, there is a clear difference between the dynamics starting from $S^z(0)=0$ and $S^z(0)=1$ rooted in the presence of the solvable eigenmodes in the GKSL equation.

\begin{figure}[t]
    \centering
    \includegraphics[width=0.45\linewidth]{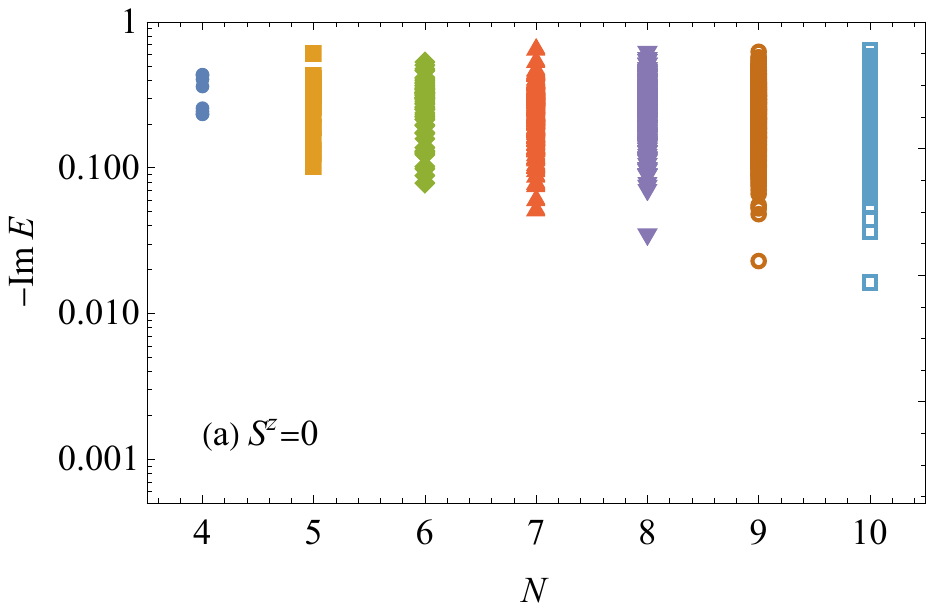}
    \includegraphics[width=0.45\linewidth]{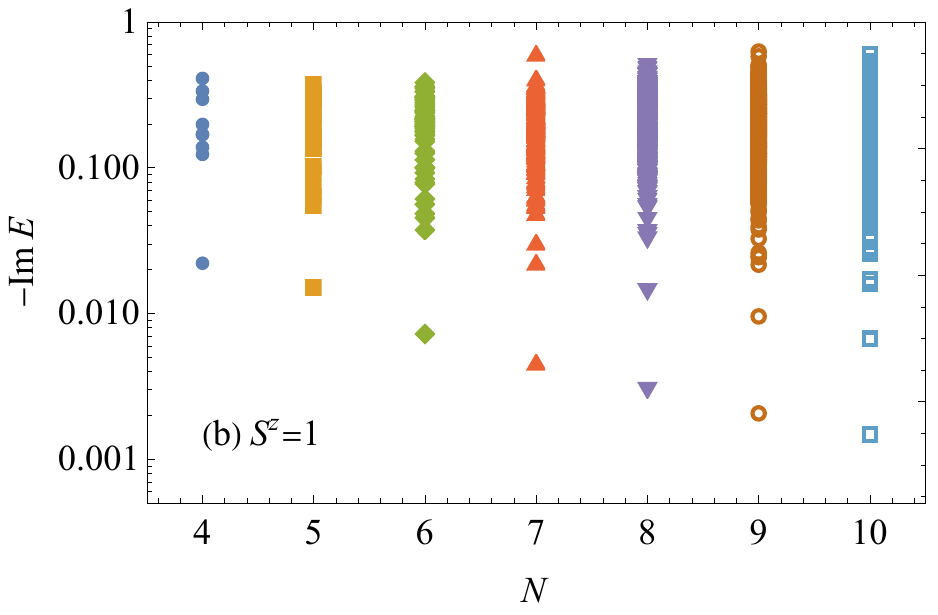}
    \caption{The imaginary part of the eigenvalues of the non-Hermitian Hamiltonian $H^{\rm eff}_{\rm AKLT}=H_{\rm AKLT}-\frac{i}{2}\sum_{\alpha={\rm L,R}} \gamma_\alpha A_\alpha^\dagger A_\alpha$ with $\gamma_{\rm L}=\gamma_{\rm R}=1$ for (a) $S^z=0$ and (b) $S^z=1$. In the case of (b) $S^z=1$, there exists one eigenvalue with a vanishing imaginary part (corresponding to the solvable dark state) for each $N$, which is not shown in the log-scale plot.}
    \label{fig:eigenvalue}
\end{figure}

The long-lived coherent oscillations observed in Fig.~\ref{fig:Sz}(b), (d) are not due to the solvable dark states, since each trajectory has a single value of $S^z$ at each time, which allows for realization of a single dark state at each time in each trajectory. One cannot have quantum mechanical superposition of different dark states,
which might cause coherent oscillations due to interference among dark states. To find the origin of the oscillations, we plot the imaginary part of the eigenvalues of the non-Hermitian Hamiltonian $H^{\rm eff}_{\rm AKLT}=H_{\rm AKLT}-\frac{i}{2}\sum_{\alpha={\rm L,R}} \gamma_\alpha A_\alpha^\dagger A_\alpha$ 
with $S^z=0$ in Fig.~\ref{fig:eigenvalue}(a) and $S^z=1$ in Fig.~\ref{fig:eigenvalue}(b). In the former case ($S^z=0$), there is no eigenstate having an eigenvalue with zero imaginary part.
All the eigenstates have nonzero imaginary parts, which are forming continuous spectra. 
In the latter case ($S^z=1$), on the other hand, there is one eigenstate having an eigenvalues with zero imaginary part, which is not shown in Fig.~\ref{fig:eigenvalue}(b) where we take a log scale in the vertical axis. This corresponds to one of the solvable eigenmodes in the GKSL equation. On top of that, we find another eigenstate having an eigenvalue with a nonzero but very small imaginary part, which is separated from those of the other eigenstates. This eigenstate has the second smallest real part of the eigenvalues. Although it is not exactly a dark state, it may create oscillations with lifetime longer than the maximum time in Fig.~\ref{fig:Sz}. In fact, we confirm that this eigenstate has a strong overlap with the initial state close to the N\'eel state for $S^z(0)=1$. If one waits for sufficiently long time, we expect that the coherent oscillations will vanish eventually.

We also performed numerical simulations for other choices of the model parameters away from the solvable regions. The results show similar behaviors (e.g., the number of trajectories with $S^z<S_{\rm max}^z-2$ decay to zero) as in the case of $S^z(0)=0$ for the AKLT model with the boundary dissipators. Again we confirm the role of the rSGA-induced solvable eigenmodes in the GKSL equation, which support nontrivial steady states in the presence of dissipations.

%
\subsection{HSF-induced solvable eigenmodes}

The second example of robust partial solvability against boundary dissipators can be found for systems with the HSF. 
The HSF of the Liouvillian has been discussed in the context of the commutant algebra~\cite{bib:MM22, bib:LSP23}. 
In this subsection, on the other hand, we discuss partial solvability induced by the HSF structure of the Hamiltonian for boundary dissipative systems, which have not been considered before. 

In order to discuss the HSF for Liouvillians, it is useful to work on the TFD vector expression~\cite{bib:I76, bib:M03}, which is realized by the isomorphism,
\begin{align} \label{eq:TFD}
    \varphi:\, 
    &\rho = \sum_{m,n} p_{m,n} |m \rangle \langle n| 
    \mapsto |\rho \rangle\!\rangle = \sum_{m,n} p_{m,n} |m \rangle \otimes |n \rangle^*. 
\end{align}
In the TFD expression, the Liouvillian is expressed as a non-Hermitian Hamiltonian acting on the doubled Hilbert space $\mathcal{H} \otimes \mathcal{H}^*$,
\begin{align} \label{eq:effective_H}
    &\frac{d}{dt} |\rho(t) \rangle\!\rangle = -i \widetilde{H} |\rho(t) \rangle\!\rangle, \\
    &\widetilde{H} = H \otimes \bm{1} - \bm{1} \otimes {^t\!}H + i \sum_{\alpha} \gamma_{\alpha} \left( (A_{\alpha} \otimes A_{\alpha}^*) - \frac{1}{2} (A_{\alpha}^{\dag} A_{\alpha} \otimes \bm{1} + \bm{1} \otimes {^t\!}A_{\alpha} A_{\alpha}^*) \right). \nonumber
\end{align}
For boundary dissipative systems, the quantum jump operators $A_{\alpha}$ non-trivially act only on the first and/or the $N$th site. In this subsection, we consider the Hamiltonian with the HSF that is solvable at least in one of the fragmented subspaces. 
%
Throughout this subsection, we consider the (perturbed) $XXC$ Hamiltonian \eqref{eq:XXC} that exhibits both the HSF and (partial) integrability. 
%
%

\begin{figure} 
\begin{center}
    \includegraphics[width = 145mm]{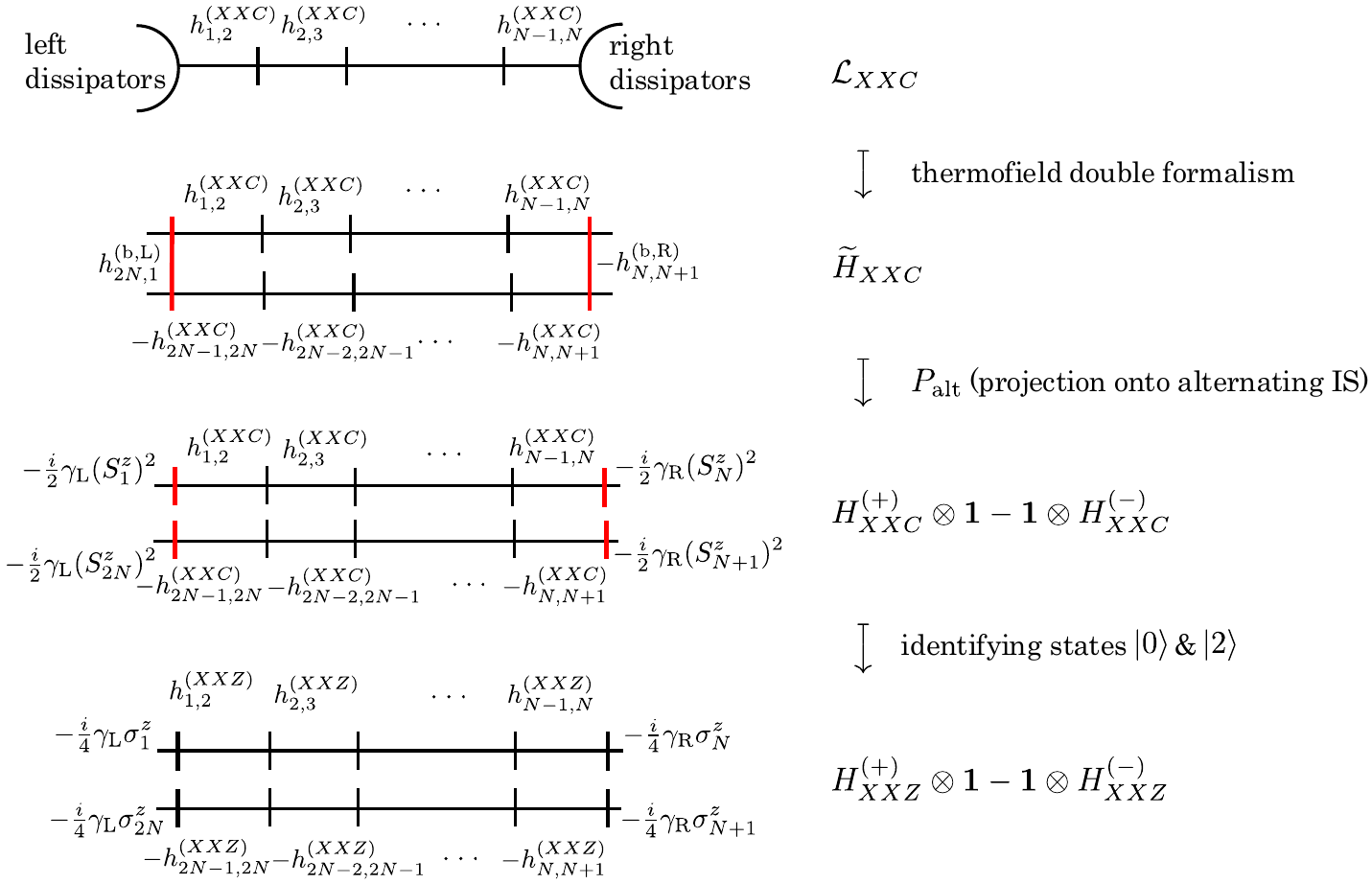}
    \caption{The $XXC$ spin chain coupled to boundary dissipators (the first line) is mapped to two $XXC$ spin chains interacting each other at the boundaries via the thermofield double formalism (the second line). In the integrable subspace specified by the alternating IS, this $XXC$ spin chain of of length $2N$ decouples into two non-interacting spin chains with imaginary boundary magnetic fields (the third line). By identifying two local states $|0 \rangle$ and $|2 \rangle$, these two decoupled $XXC$ spin chains are mapped to two $XXZ$ spin chains under boundary magnetic fields in the $z$-direction (the fourth line), which are known to be integrable. } \label{fig:open_XXC}
\end{center}
\end{figure}

Suppose that two kinds of dissipators are coupled to each end of the $XXC$ spin chain,
\begin{align} \label{eq:dissipators}
    A_{{\rm L},+} = (S_1^+)^2, \quad A_{{\rm L},-} = (S_1^-)^2, \quad 
    A_{{\rm R},+} = (S_N^+)^2, \quad A_{{\rm R},-} = (S_N^-)^2,    
\end{align}
with the coupling strengths controlled by the dissipation rates $\gamma_{{\rm L},+}$, $\gamma_{{\rm L},-}$, $\gamma_{{\rm R},+}$, and $\gamma_{{\rm R},-}$. 
With these dissipators, the Liouviilian for the boundary dissipative $XXC$ spin chain $\mathcal{L}_{XXC}$ is effectively written as the spin chain having twice the length $2N$ than the original chain (Fig.~\ref{fig:open_XXC}),
\begin{align} \label{eq:non_Hermitian}
    &\widetilde{H}_{XXC} = \sum_{j=1}^{N-1} h^{(XXC)}_{j,j+1} + h^{({\rm b,R})}_{N,N+1} 
    - \sum_{j=N+1}^{2N-1} h^{(XXC)}_{j,j+1} + h^{({\rm b,L})}_{1,2N}.
\end{align}
Here we have used the transpose invariance and the inversion symmetry of the $XXC$ Hamiltonian \eqref{eq:XXC},
\begin{align}
    &{^t}\Big(\sum_{j=1}^{N-1} h_{j,j+1}^{XXC} \Big) = \sum_{j=1}^{N-1} {^t}h_{j,j+1}^{XXC} = \sum_{j=1}^{N-1} h_{j,j+1}^{XXC}, \\
    &U_{\rm I} \Big(\sum_{j=1}^{N-1} h_{j,j+1}^{XXC} \Big) U_{\rm I} 
    = \sum_{j=1}^{N-1} U_{\rm I} h_{j,j+1}^{XXC} U_{\rm I}
    = \sum_{j=1}^{N-1} h_{N-j+1,N-j}^{XXC} = \sum_{j=1}^{N-1} h_{j,j+1}^{XXC}, \nonumber
\end{align}
in which the operator $U_{\rm I}$ reflects the spin chain with respect to its center,
\begin{align}
    U_{\rm I}:\, \mathcal{H} \to \mathcal{H},\quad |s_1,s_2,\dots,s_N \rangle \mapsto |s_N,\dots,s_2,s_1 \rangle. 
\end{align}
The effects of the boundary dissipators can be written in terms of interactions between the first and $2N$th sites and between the $N$th and $(N+1)$th sites,
\begin{align} \label{eq:dissipation_terms}
    h^{(\rm b,\alpha)} &= i\gamma_{{\rm \alpha},+} \Big( |00 \rangle \langle 22| - \frac{1}{2} (|2 \rangle \langle 2| \otimes \bm{1} + \bm{1} \otimes |2 \rangle \langle 2|) \Big) \\
    &\quad + i\gamma_{{\rm \alpha},-} \Big( |22 \rangle \langle 00| - \frac{1}{2} (|0 \rangle \langle 0| \otimes \bm{1} + \bm{1} \otimes |0 \rangle \langle 0|) \Big), \quad
    \alpha \in \{{\rm R,L}\}, \nonumber
\end{align}
each of which represents incoming and outgoing quasiparticles.

When all the dissipation rates are set to zero, the effective Hamiltonian \eqref{eq:non_Hermitian} simply consists of the two decoupled $XXC$ Hamiltonians. Therefore, the HSF structure is obtained in the doubled Hilbert space according to the IS consisting of configurations of $0$s and $2$s. Moreover, this doubled $XXC$ Hamiltonian is obviously integrable, since each of the $XXC$ spin chains is integrable. 

On the other hand, both the HSF structure and entire integrability are broken by the presence of boundary dissipation terms. However, one may notice that the subspace specified by the alternating IS survives as an invariant subspace of the entire effective Hamiltonian \eqref{eq:non_Hermitian} even in the presence of the boundary dissipation terms, since the nearest-neighbor interactions in the boundary dissipation terms \eqref{eq:dissipation_terms}, i.e. the IS violating terms, are irrelevant in this subspace. 
In the original Liouvillian expression, this means that any state with the alternating IS never goes out from the other subspaces, nor any state in the other subspaces never come into the subspace with the alternating IS.

%
\subsubsection{Integrable subspaces}
The diagonal terms in \eqref{eq:dissipation_terms} can be regarded as boundary magnetic fields on each of the two $XXC$ Hamiltonians, and therefore the subspace specified by the alternating IS is the integrable subspace of the effective Hamiltonian \eqref{eq:non_Hermitian} if these diagonal terms provide the integrable boundaries of the $XXC$ model. 

The integrable boundary conditions for the $XXC$ model have been discussed in \cite{bib:AM98}. 
When boundary magnetic fields are imposed only in the $z$-direction, the Hamiltonian is integrable if the boundary terms are given by 
\begin{align}
    H_{{\rm b}XXC} &= \sum_{j=1}^{N-1}h^{XXC}_{j,j+1} 
    + \sinh\eta \coth\xi_- \cdot (S_1^z)^2
    - \sinh\eta \coth\xi_+ \cdot (S_N^z)^2 \\
    &\quad + \frac{1}{2} \sinh\eta (\coth\xi_+ - \coth\xi_-).  \nonumber
\end{align}
Here $\xi_-$ and $\xi_-$ are arbitrary complex parameters. 
%
With these observations, the effective Hamiltonian \eqref{eq:non_Hermitian} becomes integrable in the subspace specified by the alternating IS when the quasiparticle incoming and outgoing rates are the same at each end,
\begin{align}
    \gamma_{\alpha,+} = \gamma_{\alpha,-}, \quad
    \alpha \in \{ {\rm L}, {\rm R} \}. 
\end{align}

Although we set $H$ as the integrable $XXC$ Hamiltonian so far, it is also possible to replace it with the perturbed $XXC$ Hamiltonian by adding site-dependent perturbations on the bulk without violating partial solvability of the Liouvillian in the subspace with the alternating IS. Such a perturbation \eqref{eq:pert_1} modifies the bulk interactions in the effective Hamiltonian \eqref{eq:non_Hermitian} as 
\begin{align} \label{eq:pert_3}
    &h^{XXC}_{j,j+1} \to h^{XXC}_{j,j+1} + h_{j,j+1}^{\rm alt}(\beta^{(j)}_{{\rm d}1},\beta^{(j)}_{{\rm d}2},\beta^{(j)}_{{\rm o}1},\beta^{(j)}_{{\rm o}2},\zeta^{(j)}), \quad j=1,\dots,N-1, \\
    &h^{XXC}_{j,j+1} \to h^{XXC}_{j,j+1} + h_{j,j+1}^{\rm alt}(\beta^{(j)}_{{\rm d}1},\beta^{(j)}_{{\rm d}2},\beta^{(j)}_{{\rm o}2},\beta^{(j)}_{{\rm o}1},\zeta^{(j)}), \quad j=N+1,\dots,2N-1, \nonumber
\end{align}
both of which are apparently in the form of \eqref{eq:pert_1}, implying that the effective Hamiltonian stays partially solvable in the subspace with the alternating IS after introducing these bulk perturbations.

\subsubsection{Solvable eigenmodes}

The $XXC$ Hamiltonian with the integrable boundaries can be diagonalized via the Bethe ansatz method due to the existence of the $R$- and $K$-matrices which solve the Yang-Baxter equation and the reflection relation, respectively. For the perturbed $XXC$ model that becomes integrable only in the subspace specified by a certain IS,
the spectrum and eigenvectors in this subspace can be found by mapping the Hamiltonian restricted in the solvable subspace to the spin-$1/2$ $XXZ$ model. 

By employing the same strategy explained in Sec.~\ref{sec:closed_HSF_pert}, the effective Hamiltonian $\widetilde{H}_{XXC}$ \eqref{eq:non_Hermitian} in the solvable subspace, namely the subspace specified by the alternating IS, is first mapped to two decoupled $XXC$ spin chains (Fig.~\ref{fig:open_XXC}). Then these $XXC$ spin chains can be mapped to the spin-$1/2$ $XXZ$ chains with the diagonal boundary magnetic fields by identifying the states $|0 \rangle$ and $|2 \rangle$,
\begin{align} \label{eq:rnon-Hermitian}
    &\widetilde{H}_{XXC} \underset{P_{\rm alt} \mathcal{H} \setminus \{ |0\rangle,|2\rangle \}^N}{\longmapsto} H^{(+)}_{XXZ}(\gamma_{\rm L},\gamma_{\rm R}) \otimes \bm{1} - \bm{1} \otimes H^{(-)}_{XXZ}(\gamma_{\rm R},\gamma_{\rm L}), \\
    &H^{(+)}_{XXZ}(\gamma_{\rm L},\gamma_{\rm R}) = - \frac{i}{4} \gamma_{\rm L} \sigma_1^z + \sum_{j=1}^{N-1} \Big( \sigma_j^+ \sigma_{j+1}^- + \sigma_j^- \sigma_{j+1}^+ + \frac{1}{2}\cosh\eta\, \sigma_j^z \sigma_{j+1}^z \Big) - \frac{i}{4} \gamma_{\rm R} \sigma_N^z \nonumber \\
    &\hspace{32mm}+ \frac{N-1}{2} \cosh\eta - \frac{i}{4} (\gamma_{\rm L} + \gamma_{\rm R}), \nonumber
    \\
    &H^{(-)}_{XXZ}(\gamma_{\rm R},\gamma_{\rm L}) = - \frac{i}{4} \gamma_{\rm R} \sigma_{N+1}^z - \sum_{j=N+1}^{2N-1} \Big( \sigma_j^+ \sigma_{j+1}^- + \sigma_j^- \sigma_{j+1}^+ + \frac{1}{2}\cosh\eta\, \sigma_j^z \sigma_{j+1}^z \Big) - \frac{i}{4} \gamma_{\rm L} \sigma_{2N}^z \nonumber \\
    &\hspace{32mm}+ \frac{N-1}{2} \cosh\eta - \frac{i}{4} (\gamma_{\rm L} + \gamma_{\rm R}).  \nonumber
\end{align}
Note that the boundary magnetic fields here are ``imaginary magnetic fields" with pure imaginary coefficients. By writing sets of energy eigenvalues for these two spin-$1/2$ $XXZ$ chains as $\{E^{(+)}_{n_+}\}_{n_+}$ and $\{E^{(-)}_{n_-}\}_{n_-}$, the set of the summed energy eigenvalues $\{E^{(+)}_{n_+} + E^{(-)}_{n_-}\}_{n_+,n_-}$ is embedded in the full spectrum of the effective Hamiltonian $\widetilde{H}_{XXC}$. 
Since the map $\varphi$ is an isomorphism, the spectrum of the Liouvillian $\mathcal{L}_{XXC}$ matches that of the effective Hamiltonian $\widetilde{H}_{XXC}$. The energy eigenvalues of the double spin chain, $\{E^{(+)}_{n_+} + E^{(-)}_{n_-}\}_{n_+,n_-}$, then agree with the eigenvalues of the Liouvillian $\mathcal{L}_{XXC}$ restricted in the solvable subspace. 

The corresponding eigenvectors $|E^{(+)}_{n_+} \rangle \otimes |E^{(-)}_{n_-} \rangle$ thus provide the eigenvectors of the effective non-Hermitian Hamiltonian $\widetilde{H}_{XXC}$ by a map 
\begin{align}
    &|\sigma_1 \sigma_2 \dots \sigma_N \rangle \in \mathbb{C}^{2N}
    \mapsto |\tau_1 \tau_2 \dots \tau_N \rangle \in \mathbb{C}^{3N}, \\
    &\tau_j = 1 + \theta_j^{\omega_j}, \nonumber
\end{align}
in which $\theta_j$ and $\omega_j$ are defined by 
\begin{align}
    &\theta_j = -|1 - \sigma_j|, \qquad
    \omega_j = 1 - (-1)^{\sum_{k=1}^j \sigma_k},
\end{align}
respectively.
Here $\theta_j$ is a parameter that determines whether the $j$th site is in the local state $|1 \rangle$ or not, while $\omega_j$ counts the number of $0$s and $2$s between the first and $j$th site. 
The eigenvectors of the effective non-Hermitian Hamiltonian $\widetilde{H}_{XXC}$ are mapped to the eigenmodes of the Liouvillian $\mathcal{L}_{XXC}$ via the inverse map of the isomorphism \eqref{eq:TFD}. 
Therefore, solving the eigenvalue problem for the spin-$1/2$ $XXZ$ chain under the imaginary boundary magnetic fields tells the eigenmodes for the spin-$1$ $XXC$ model coupled to boundary dissipators. 


The Bethe-ansatz method is well established for the spin-$1/2$ $XXZ$ chain even in the presence of boundary magnetic fields~\cite{bib:S88}. The Hamiltonian and conserved quantities are constructed from a series expansion of the transfer matrix, which consists of the $R$-matrix
\begin{align}
    R_{ij}(\lambda) &= 
    \sinh\left(\lambda+ \frac{\eta}{2}\right) \cosh\frac{\eta}{2} \cdot \bm{1}_{ij} + \cosh\left(u+\frac{\eta}{2}\right) \sinh\frac{\eta}{2} \cdot \sigma_i^z \sigma_j^z \\
    &\quad + \sinh\eta \cdot \left( \sigma_i^+ \sigma_j^- + \sigma_i^- \sigma_j^+ \right), \nonumber
\end{align}
and the $K$-matrix
\begin{align} \label{eq:K-matrix}
    K(\lambda,\xi) = \sinh\xi \cosh\lambda \cdot \bm{1} + \cosh\xi \sinh\lambda \cdot \sigma^z, 
\end{align}
which solve the Yang-Baxter equation \eqref{eq:YBE} and the reflection relation,
\begin{align} \label{eq:reflection}
    R_{12}(\lambda_1-\lambda_2) K_1(\lambda_1) R_{12}(\lambda_1+\lambda_2) K_2(\lambda_2)
    = K_2(\lambda_2) R_{12}(\lambda_1 + \lambda_2) K_1(\lambda_1) R_{12}(\lambda_1-\lambda_2). 
\end{align}
The complex parameter $\xi$ determines the strength of the diagonal boundary magnetic fields, which apprears in the spin-$1/2$ $XXZ$ Hamiltonian as  
\begin{align} \label{eq:XXZ}
    H_{XXZ} 
    &= \sum_{j=1}^{N-1} \left( \sigma_j^+ \sigma_{j+1}^- + \sigma_j^- \sigma_{j+1}^+ + \frac{1}{2}\cosh\eta \cdot \sigma_j^z \sigma_{j+1}^z \right) \\
    &\quad +\frac{1}{2} \sinh\eta \coth\xi_- \cdot \sigma_1^z  - \frac{1}{2} \sinh\eta \coth\xi_+ \cdot \sigma_N^z. \nonumber
\end{align}
Thus, in order to realize the restricted effective Hamiltonian \eqref{eq:rnon-Hermitian},
we need to choose $\xi_{\pm}$ to be pure imaginary for $\eta \in \mathbb{R}$ (i.e., in the gapped regime) or 
to be real for pure imaginary $\eta$ (i.e., in the gapless regime).

The transfer matrix $T(\lambda)$ for the open spin chain is ``a double-row transfer matrix", which consists of two products of the $R$-matrices,
\begin{align}
    &T(\lambda) = {\rm tr}_0 \left( K_0\left(-\lambda - \eta,\xi_+\right)\, M_0(\lambda)\, K_0\left(\lambda,\xi_-\right)\, \widehat{M}_0(\lambda) \right), \\
    &M_0(\lambda) = R_{0N}(\lambda) \dots R_{01}(\lambda), \nonumber \\
    &\widehat{M}_0(\lambda) = R_{10}(\lambda) \dots R_{N0}(\lambda) \nonumber. 
\end{align}
These transfer matrices are mutually commuting,
\begin{align}
    [T(\lambda),\, T(\mu)] = 0,
\end{align}
for any $\lambda, \mu \in \mathbb{C}$. 
Thus, a series expansion of the transfer matrix,
\begin{align}
    T(\lambda) = \exp\left( \sum_r \frac{\lambda^r}{r!} Q_r \right),
\end{align}
provides a large number of conserved quantities $Q_r$. By cumbersome but straightforward calculations, one can confirm that the $XXZ$ Hamiltonian with the boundary magnetic fields \eqref{eq:XXZ} is obtained 
from $Q_1$ up to a constant~\cite{bib:S88, bib:N04},
\begin{align}
    H_{XXZ} &= \left(2 \sinh\xi_+ \sinh\xi_- \cosh\eta\, (\sinh\eta)^{2N-1}\right)^{-1} \cdot \frac{d}{d\lambda} T(\lambda) \Big|_{\lambda = 0}
    - \frac{(\sinh\eta)^2 + N (\cosh\eta)^2}{\cosh\eta}. 
\end{align}

The eigenvectors of the Hamiltonian are derived by diagonalizing the transfer matrix, which is achieved by the Bethe-ansatz method~\cite{bib:N04}. The eigenenergies are written in terms of the eigenvalues of the transfer matrix $\tau(\lambda)$,
\begin{align} \label{eq:eigenenergy}
    &E(\{\lambda_j\}) = \left(2 \sinh\xi_+ \sinh\xi_- \cosh\eta\, (\sinh\eta)^{2N-1}\right)^{-1} \cdot \tau'(0) - \frac{(\sinh\eta)^2 + N (\cosh\eta)^2}{\cosh\eta}, \\
    &\tau(\lambda) = (\sinh(\lambda + \eta))^{2N} \frac{\sinh(2\lambda + 2\eta)}{\sinh(2\lambda + \eta)} \sinh(\lambda + \xi_+) \sinh(\lambda + \xi_-) \prod_{i=1}^n \frac{\sinh(\lambda - \lambda_i - \eta)}{\sinh(\lambda + \lambda_i + \eta)} \nonumber \\
    &\hspace{8mm}
    + (\sinh \lambda)^{2N} \frac{\sinh(2\lambda)}{\sinh(2\lambda + 
\eta)} \sinh(\lambda + \eta - \xi_+) \sinh(\lambda + \eta - \xi_-) \prod_{i=1}^n \frac{\sinh(\lambda + \lambda_i + 2\eta)}{\sinh(\lambda - \lambda_i)}, \nonumber
\end{align}
where $\lambda_i$ solves a set of the Bethe equations,
\begin{align} \label{eq:Bethe_eq}
    \left( \frac{\sinh(\lambda_j + \eta)}{\sinh(\lambda_j)} \right)^{2N}
    &= \frac{\sinh(\lambda_j - \xi_+ + \eta) \sinh(\lambda_j - \xi_- + \eta)}{\sinh(\lambda_j + \xi_+) \sinh(\lambda_j + \xi_-)} \\
    &\quad\cdot \prod_{\genfrac{}{}{0pt}{}{k=1}{k \neq j}}^n 
    \frac{\sinh(\lambda_j - \lambda_k + \eta)}{\sinh(\lambda_j - \lambda_k - \eta)} \frac{\sinh(\lambda_j + \lambda_k + 2\eta)}{\sinh(\lambda_j + \lambda_k)}, \nonumber
\end{align}
for $j = 1,2,\dots,n$. 

\begin{figure} 
\begin{center}
    \includegraphics[width = 95mm]{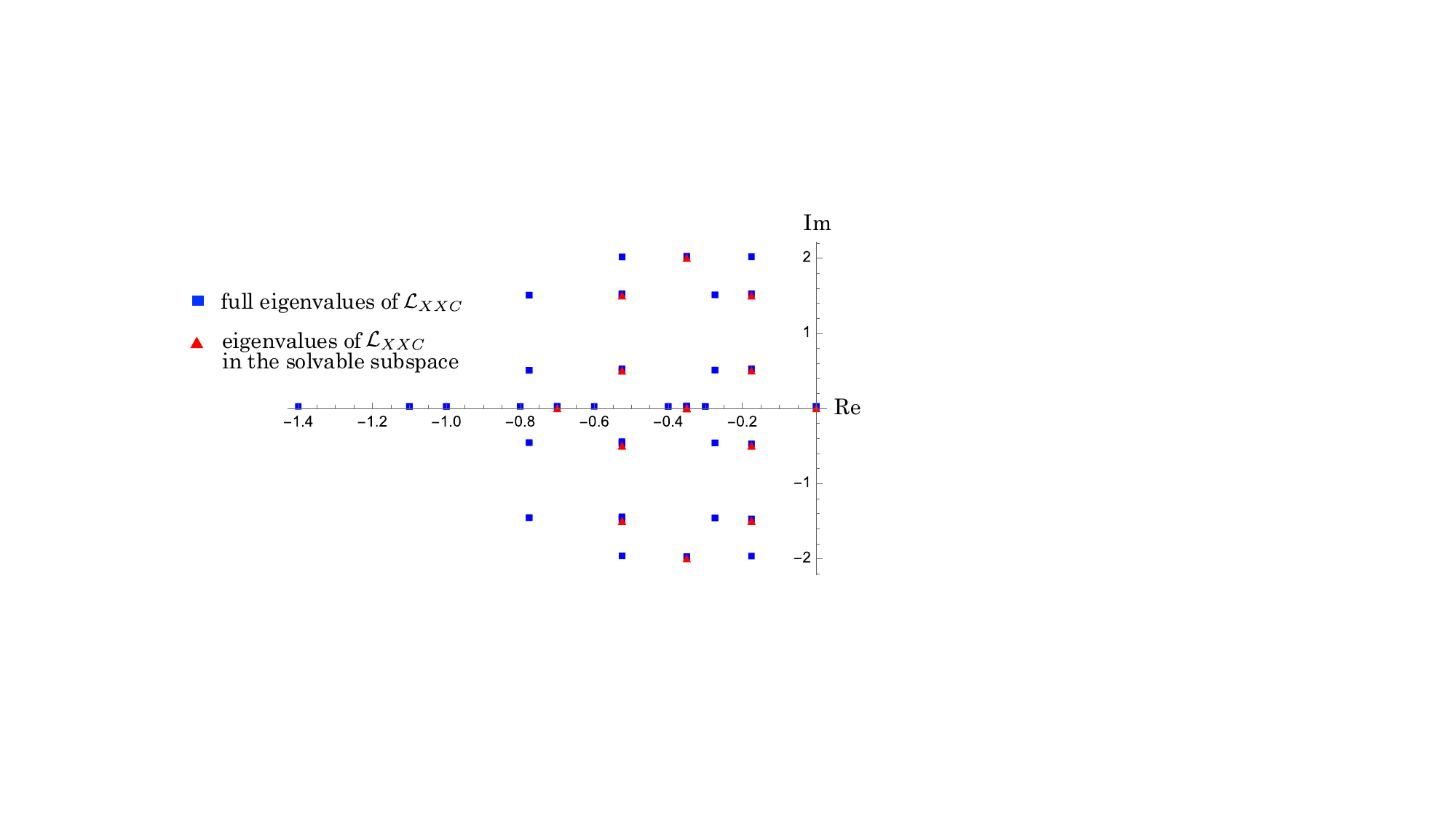}
    \caption{The spectra of the Liouvillian $\mathcal{L}_{XXC}$ in the entire Hilbert space and integrable subspace are plotted for the $N = 2$ $XXC$ chain coupled to the boundary dissipators. The anisotropy and the dissipation rates are set as $\eta = i\pi/3$, $\gamma_{\rm L} = 0.4$, and $\gamma_{\rm R} = 0.3$. The full spectrum for $\mathcal{L}_{XXC}$ is plotted by the square dots, while the spectrum of $\mathcal{L}_{XXC}$ in the integrable subspace is plotted by the triangle dots. The latter spectrum is fully included in the former spectrum. } \label{fig:spectrum_L}
\end{center}
\end{figure}

Since the analytic solutions to the Bethe equations \eqref{eq:Bethe_eq} are inaccessible, we instead give numerical results for the energy spectra of the $XXZ$ spin chains (Fig.~\ref{fig:spectrum_L}). As compared from the spectrum of the effective Hamiltonian \eqref{eq:non_Hermitian}, the sums of the eigenenergies for $H^{(+)}$ and $H^{(-)}$ are indeed embedded in its full spectrum. 
Unlike the rSGA-induced partial solvability discussed in the previous subsection, neither equally-spaced spectrum nor pure-imaginary eigenvalue is observed in the solvable subspace of the $XXC$ model.
This indicates that no oscillating mode exists for the $XXC$ model coupled to the boundary dissipators. The only non-decaying mode is the steady state, corresponding to the zero eigenvalue of the effective Hamiltonian. 
Degenerate steady states are observed for the entire Liouvillian, but the integrable subspace has the unique steady state,
\begin{align}
    \rho_{\rm ss}^{XXC} = |11 \dots 1 \rangle \langle 11 \dots 1|,    
\end{align}
which is a product state. This is also a completely separated state from the other states by the HSF, and therefore initial states never reach the integrable steady state unless it is the steady state itself.

\subsubsection{Solvable steady states induced by MPO symmetry}
In the previous subsection, we have seen that partial solvability emerges for the open $XXC$ model coupled to boundary dissipators, where the eigenmodes including the steady state are exactly calculated. 
If one focuses only on steady states, some of them are analytically derived even if they are out of the solvable subspace. 

The derivation of those ``out-of-integrable" steady states can be achieved based on the MPO symmetry we reviewed in Sec.~\ref{sec:MPO}. 
By following the method originally introduced in \cite{bib:BPP23}, we write the steady state in the form of a product of an amplitude operator $\Omega$,
\begin{align} \label{eq:SS_MPO}
    \rho_{\rm ss} = \Omega \Omega^{\dag}.
\end{align}
Then, if the amplitude operator $\Omega$ is given by the matrix product form,
\begin{align} \label{eq:amp_MPO}
    &\Omega = {_a}\langle v_{\rm L}| L_{a,N} \dots L_{a,2} L_{a,1} |v_{\rm R} \rangle_{a},
\end{align}
in which the operator $L_{a,n}$ satisfies the local divergence \eqref{eq:local_divergence} with the local perturbed $XXC$ Hamiltonian, the amplitude matrix \eqref{eq:amp_MPO} produces the steady-state density matrix, as we see in the following. 

Let us consider the time evolution of the density matrix consisting of the amplitudes matrix \eqref{eq:amp_MPO}. Time evolution of any density matrix is described by the Lindblad equation \eqref{eq:Lindblad}. Due to the local divergence relation \eqref{eq:local_divergence}, the commutator term produces only two terms coming from the non-commutativity at the boundaries,
\begin{align} \label{eq:non-commuting}
    [H_{XXC},\,\Omega] &= {_a}\langle v_{\rm L}| M_{a,N} \dots L_{a,2} L_{a,1} |v_{\rm R} \rangle_{a} 
    - {_a}\langle v_{\rm L}| L_{a,N} \dots L_{a,2} M_{a,1} |v_{\rm R} \rangle_{a} \\
    &= 0, \nonumber
\end{align}
since the solution exists only for $M = 0$. 
On the other hand, the boundary dissipators non-trivially act only at the edges, which may cancel the non-commuting terms above,
\begin{align} \label{eq:b-cancel}
    &{_{a}}\langle v_{\rm L}| \otimes {_b}\langle v_{\rm L}| \left( -i M_{a,N} L_{b,N}^{\dag_{\rm p}} + \gamma_{{\rm R},+} \mathcal{D}_{{\rm R},+}(L_{a,N} L_{b,N}^{\dag_{\rm p}}) + \gamma_{{\rm R},-} \mathcal{D}_{{\rm R},-}(L_{a,N} L_{b,N}^{\dag_{\rm p}}) \right) = 0  , \nonumber, \\
    &\left( i M_{a,1} L_{b,1}^{\dag_{\rm p}} + \gamma_{{\rm L},+} \mathcal{D}_{{\rm L},+}(L_{a,1} L_{b,1}^{\dag_{\rm p}}) + \gamma_{{\rm L},-} \mathcal{D}_{{\rm L},-}(L_{a,1} L_{b,1}^{\dag_{\rm p}}) \right) |v_{\rm R} \rangle_a \otimes |v_{\rm R} \rangle_{b} = 0. \nonumber
\end{align}
Here we leave the boundary dissipation rates free, which shall be constrained later for solvability of the steady state. 
The solutions to the divergence relation \eqref{eq:non-commuting} and the boundary cancellation condition \eqref{eq:b-cancel} then provide the steady state density matrix in the form \eqref{eq:SS_MPO} and \eqref{eq:amp_MPO}. 

We found that both the diagonal and off-diagonal MPO symmetries on the bulk result in the same solution,
\begin{align}
    &y=0, \\
    &|x|^2 = \frac{\gamma_{{\rm R},+}}{\gamma_{{\rm R},-}} |u|^2, \qquad
    |x|^2 = \frac{\gamma_{{\rm R},+}}{\gamma_{{\rm R},-}} |v|^2, \nonumber \\
    &\frac{\gamma_{{\rm R},+}}{\gamma_{{\rm R},-}} = \frac{\gamma_{{\rm L},+}}{\gamma_{{\rm L},-}} = \omega. \nonumber 
\end{align}
The integrable steady state observed in the previous subsection is realized as a special case with $u = v = 0$ and $\omega = 1$. 

\section{Concluding remarks}
\label{sec: conclusion}

In this paper, we have introduced a new class of partially solvable open quantum systems. The models are new in the sense that their partial solvability is induced by partial solvability of the Hamiltonians. The systems are coupled to dissipators only at the boundaries, unlike other existing partially solvable models~\cite{bib:P11, bib:TMBJ20, bib:WM24}, in which the dissipators are attached at every site. 
We showed that partial solvability of the system can be robust against the boundary dissipators under two different kinds of conditions. 

Liouvillians in the first type admit the existence of dark states, i.e., the states which do not feel the effect of dissipators, and hence partial solvability of the Hamiltonian survives in the subspace spanned by these dark states. A pure dark state becomes an exactly solvable steady state of the Liouvillians, while an off-diagonal element density matrix consisting of the dark states provides an eigenmode of the Liouvillian. 
As an example, we especially focused on the AKLT-type model coupled to quasiparticle baths at both edges. As a known fact, the AKLT-type model exhibits four degenerate ground states~\cite{bib:K90} due to the boundary spin fractionalization, and a tower of solvable quasiparticle excited states can be constructed on top of each ground state~\cite{bib:MRBR18, bib:CIKM23}. Among these four towers of the degenerate solvable energy eigenstates, we found that a tower of states under a certain choice of the boundary spins is a set of dark states of the Liouvillian. 
One of the remarkable features of this model is persistent oscillations obtained in local observables, when the initial state is prepared to have a large enough overlap with the solvable state of the Liouvillian. This quantum synchronization is brought by the equally-spaced spectrum of the Liouvillian in the solvable subspace consiting of the dark states, which is inherited by the equally-spaced spectrum of the Hamiltonian due to the rSGA structure. A similar phenomenon has been reported in \cite{bib:TMBJ20, bib:WM24} for the Liouvillian which also exhibits the rSGA but whose dissipators are coupled to all the sites of the system. 

The second mechanism which makes Liouvillians partially solvable is the HSF. We are especially interested in the case where the Hamiltonian exhibits the HSF which divides the Hilbert space into exponentially-many subspaces, some of which may survive even if the boundary dissipators are introduced. The key property in extending the notion of partial solvability induced by the HSF is the robustness of some subspaces under site-dependent perturbations. 
We have considered the $XXC$ spin chain as an example, which exhibits the HSF due to the presence of IS. We showed that the effect of the boundary quasiparticle baths, which inject and absorb quasiparticles, can be regarded as the partial solvability preserving perturbations through the thermofield double formalism. 
As a result, the effective Hamiltonian in the solvable subspace of the doubled Hilbert space becomes two decouples integrable $XXZ$ spin chains with imaginary boundary magnetic fields corresponding to the boundary dissipators. Thus, any eigenmode of the Liouviilian in this solvable subspace is accessible via the Bethe-ansatz method, which is numerically verified (Fig.~\ref{fig:spectrum_L}). 

Besides the eigenmodes in the solvable subspace, several more steady states can be exactly derived by using the MPO symmetry of the $XXC$ Hamiltonian, which is characterized by the local divergence relation \eqref{eq:local_divergence}. As the local divergence relation \eqref{eq:local_divergence} holds for the $XXC$ model only when it is reduced to the frustration-free condition ($M=0$ in Eq.~\eqref{eq:local_divergence}), the steady states associated with the MPO symmetry are the dark states, which vanish under the action of the dissipators. We found that the steady state in the solvable subspace is included in the class of solvable steady states associated with the MPO symmetry. 

A new class of partially solvable open quantum systems introduced in this paper paves a way to study non-integrable open quantum systems. At the same time, it also proposes several interesting questions. We list possible future works in the following, by focusing on the HSF-induced partially solvable open quantum systems.
The first question is how large one can make an overlap with the solvable subspace of the Liouvillian, when a ``physical" initial state, such as the dimer state and N\'{e}el state, is prepared.
This would be addressed by following the method introduced in \cite{bib:P18}, which allows the overlap between the initial state and the energy eigenstates to be expressed by the Tsuchiya determinant, after the thermofield double formalism is applied. 
The second question we are interested in is how the relaxation time differs between the solvable subspace and unsolvable subspaces. Definitely, the Liouvillian restricted in the solvable subspace has a gapless spectrum, as it matches the energy spectrum of the $XXZ$ model, and this may lead to a non-trivial (non-exponential) relaxation behavior. 
The third question is whether the Kardar-Parisi-Zhang (KPZ) universality class is observed for open quantum systems. As the $XXC$ chain coupled with boundary dissipators is mapped to the two $XXZ$ spin chains in the solvable subspace, it is likely that the KPZ universality class observed for the $XXZ$ spin chain~\cite{bib:LZP19} emerges also for this open quantum system, if it is robust against the boundary conditions.

\section*{Acknowledgements}
C. M. thanks H. Katsura, C. Paletta, and B. Pozsgay for helpful discussions, on which the idea of the HSF part of this work is mainly based. C. M. acknowledges financial support from JSPS KAKENHI Grant Number JP23K03244. 
N. T. acknowledges support from JST FOREST (Grant No. JP-MJFR2131) 
and JSPS KAKENHI (Grant No. JP24H00191).

\appendix
\section{Proof of Eq.~\eqref{eq:dark_state2}} \label{sec:proof}
In this Appendix, we show that any state in the subspace $W^{(\uparrow,\downarrow)}$ satisfies the dark-state condition \eqref{eq:dark_state2} in the presence of the boundary dissipators. 

As the dissipators \eqref{eq:diss_AKLT} non-trivially act only on the first and/or the $N$th site, the dark-state condition in the present case is written as 
\begin{align} \label{eq:dark_state3}
    &{_a}\langle v_{\rm L}| \otimes {_b}\langle v_{\rm L}| \mathcal{D}_{\rm L}(\vec{A} \otimes \vec{A}^{\dag_p}) = 0, \\
    &\mathcal{D}_{\rm R}(\vec{A} \otimes \vec{A}^{\dag_p}) |v_{\rm R} \rangle_a \otimes |v_{\rm R} \rangle_b = 0, \nonumber
\end{align}
in which $\vec{A} \otimes \vec{A}^{\dag_p}$ is the three-by-three matrix with the matrix-valued elements,
\begin{align} 
    \vec{A} \otimes \vec{A}^{\dag_p} = 
    \begin{pmatrix}
    A_0 \otimes A_0^* & A_0 \otimes A_1^* & A_0 \otimes A_2^* \\
    A_1 \otimes A_0^* & A_1 \otimes A_1^* & A_1 \otimes A_2^* \\
    A_2 \otimes A_0^* & A_2 \otimes A_1^* & A_2 \otimes A_2^* 
    \end{pmatrix}. 
\end{align}
The definitions of the matrices $A_0$, $A_1$, and $A_2$ are given in \eqref{eq:MPS_elements}. 

Then we have 
\begin{align}
    &\mathcal{D}_{\rm L}(\vec{A} \otimes \vec{A}^{\dag_p}) = 
    \begin{pmatrix}
    A_2 \otimes A_2^* & 0 & -\frac{1}{2} A_0 \otimes A_2^* \\
    0 & 0 & -\frac{1}{2} A_1 \otimes A_2^* \\
    -\frac{1}{2} A_2 \otimes A_0^* & -\frac{1}{2} A_2 \otimes A_1^* & A_2 \otimes A_2^*
    \end{pmatrix}, 
\end{align}
which indicates that the dissipation terms always include the element $A_2$ proportional to $\sigma^-$. Thus, the dark-state condition \eqref{eq:dark_state3} is satisfied by choosing the boundary vectors as 
\begin{align}
    &{_a}\langle v_{\rm L}| = {_a}\langle \uparrow|, \quad
    {_b}\langle v_{\rm L}| = {_b}\langle \uparrow|, \\
    &|v_{\rm R} \rangle_a = |\downarrow \rangle_a, \quad
    |v_{\rm R} \rangle_b = |\downarrow \rangle_b. \nonumber
\end{align}

\bibliographystyle{abbrv}
\bibliography{references}

\begin{thebibliography}{10}

\bibitem{bib:AKLT87}
I.~Affleck, T.~Kennedy, E.~H. Lieb, and H.~Tasaki.
\newblock {Rigorous results on valence-bond ground states in antiferromagnets}.
\newblock {\em Phys. Rev. Lett.}, 59:799, 1987.

\bibitem{bib:AAW20}
{\'{A}}.~M. Alhambra, A.~Anshu, and H.~Wilming.
\newblock {Revivals imply quantum many-body scars}.
\newblock {\em Phys. Rev. B}, 101:205107, 2020.

\bibitem{bib:AM98}
D.~Arnaudon and Z.~Maassarani.
\newblock {Integrable open boundary conditions for XXC models}.
\newblock {\em JHEP}, 10:024, 1998.

\bibitem{bib:A89}
D.~P. Arovas.
\newblock {Two exact excited states for the $S = 1$ AKLT chain}.
\newblock {\em Phys. Lett. A}, 137:431, 1989.

\bibitem{bib:BD94}
M.~Barma and D.~Dhar.
\newblock {Slow Relaxation in a Model with Many Conservation Laws: Deposition
  and Evaporation of Trimers on a Line}.
\newblock {\em Phys. Rev. Lett.}, 73:2135, 1994.

\bibitem{bib:BBN88}
A.~Barut, A.~B{\"{o}}hm, and Y.~Ne'eman.
\newblock {\em {Dynamical Groups and Spectrum Generating Algebras Vol. 1}}.
\newblock World Scientific, 1988.

\bibitem{bib:BB65}
A.~O. Barut and A.~B{\"{o}}hm.
\newblock {Dynamical Groups and Mass Formula}.
\newblock {\em Phys. Rev.}, 139:B1107, 1965.

\bibitem{bib:BO00}
C.~D. Batista and G.~Ortiz.
\newblock {Quantum Phase Diagram of the $t$-$J_z$ Chain Model}.
\newblock {\em Phys. Rev. Lett.}, 85:4755, 2000.

\bibitem{bib:B2017}
H.~Bernien, S.~Schwartz, A.~Keesling, H.~Levine, A.~Omran, H.~Pichler, S.~Choi,
  A.~S. Zibrov, M.~Endres, M.~Greiner, V.~Vuleti{\'{c}}, and M.~D. Lukin.
\newblock {Probing many-body dynamics on a 51-atom quantum simulator}.
\newblock {\em Nature}, 551:579, 2017.

\bibitem{bib:B31}
H.~Bethe.
\newblock {Zur Theorie der Metalle. I. Eigenwerte und Eigenfunktionen der
  linearen Atomkette}.
\newblock {\em Zeit. Phys.}, 71:205, 1931.

\bibitem{bib:BPP23}
M.~Borsi, L.~Pristy{\'{a}}k, and B.~Pozsgay.
\newblock {Matrix Product Symmetries and Breakdown of Thermalization from Hard
  Rod Deformations}.
\newblock {\em Phys. Rev. X}, 131:037101, 2023.

\bibitem{bib:BP07}
H.~P. Breuer and F.~Petruccione.
\newblock {\em {The Theory of Open Quantum Systems}}.
\newblock Oxford University Press, 2007.

\bibitem{bib:BBMJ20}
B.~Bu{\v{c}}a, C.~Booker, M.~Medenjak, and D.~Jaksch.
\newblock Bethe ansatz approach for dissipation: exact solutions of quantum
  many-body dynamics under loss.
\newblock {\em New J. Phys.}, 22:123040, 2020.

\bibitem{bib:CYSW13}
J.~Cao, W.~L. Yang, K.~Shi, and Y.~Wang.
\newblock {Off-Diagonal Bethe Ansatz and Exact Solution of a Topological Spin
  Ring}.
\newblock {\em Phys. Rev. Lett.}, 111:137201, 2013.

\bibitem{bib:Carmichael93}
H.~Carmichael.
\newblock {\em An Open Systems Approach to Quantum Optics}.
\newblock Springer, Berline, 1993.

\bibitem{bib:CIKM23}
A.~Chandran, T.~Iadecola, V.~Khemani, and R.~Moessner.
\newblock {Quantum Many-Body Scars: A Quasiparticle Perspective}.
\newblock {\em ANN. REV. COND. MAT. PHYS.}, 14:443, 2023.

\bibitem{bib:CC21}
E.~Chertkov and B.~K. Clark.
\newblock {Motif magnetism and quantum many-body scars}.
\newblock {\em Phys. Rev. B}, 104:104410, 2021.

\bibitem{bib:Daley14}
A.~J. Daley.
\newblock Quantum trajectories and open many-body quantum systems.
\newblock {\em Advances in Physics}, 63(2):77--149, 2014.

\bibitem{bib:DCM92}
J.~Dalibard, Y.~Castin, and K.~M{\o}lmer.
\newblock Wave-function approach to dissipative processes in quantum optics.
\newblock {\em Phys. Rev. Lett.}, 68:580--583, 1992.

\bibitem{bib:DFJMN93}
B.~Davies, O.~Foda, M.~Jimbo, T.~Miwa, and A.~Nakayashiki.
\newblock {Diagonalization of the $XXZ$ Hamiltonian by vertex operators}.
\newblock {\em Commun. Math. Phys.}, 151:89, 1993.

\bibitem{bib:THSP19}
G.~{D}e Tomasi, D.~Hetterich, P.~Sala, and F.~Pollmann.
\newblock {Dynamics of strongly interacting systems: From Fock-space
  fragmentation to many-body localization}.
\newblock {\em Phys. Rev. B}, 100:214313, 2019.

\bibitem{bib:DHTP21}
J.~Y. Desaules, A.~Hudomal, C.~J. Turner, and Z.~Papi{\'{c}}.
\newblock {Proposal for Realizing Quantum Scars in the Tilted 1D Fermi-Hubbard
  Model}.
\newblock {\em Phys. Rev. Lett.}, 126:210601, 2021.

\bibitem{bib:DB93}
D.~Dhar and M.~Barma.
\newblock {Conservation laws in stochastic deposition-evaporation models in one
  dimension}.
\newblock {\em Pramana}, 41:L193, 1993.

\bibitem{bib:DMKKBZ08}
S.~Diehl, A.~Micheli, A.~Kantian, B.~Kraus, H.~P. B{\"{u}}chler, and P.~Zoller.
\newblock {Quantum states and phases in driven open quantum systems with cold
  atoms}.
\newblock {\em Nature Phys.}, 4:878, 2008.

\bibitem{bib:DGN65}
Y.~Dothan, M.~Gell-Mann, and Y.~Ne’eman.
\newblock {Series of hadron energy levels as representations of non-compact
  groups}.
\newblock {\em Phys. Lett.}, 17:148, 1965.

\bibitem{bib:EP20}
F.~H.~L. Essler and L.~Piroli.
\newblock {Integrability of one-dimensional Lindbladians from operator-space
  fragmentation}.
\newblock {\em Phys. Rev. E}, 102:062210, 2020.

\bibitem{bib:F96}
L.~D. Faddeev.
\newblock {How Algebraic Bethe Ansatz works for integrable model}.
\newblock arXiv:hep-th/9605187, 1996.

\bibitem{bib:FT84}
L.~D. Faddeev and L.~A. Takhtadzhyan.
\newblock {Spectrum and scattering of excitations in the one-dimensional
  isotropic Heisenberg model}.
\newblock {\em J. Sov. Math.}, 24:241, 1984.

\bibitem{bib:BO93}
B.~Gruber and T.~Otsuka.
\newblock {\em {Symmetries in science VII: spectrum-generating algebras and
  dynamic symmetries in physics}}.
\newblock Springer, 1993.

\bibitem{bib:HCLYSW14}
K.~Hao, J.~Cao, G.~L. Li, W.~L. Yang, K.~Shi, and Y.~Wang.
\newblock {Exact solution of the Izergin-Korepin model with general
  non-diagonal boundary terms}.
\newblock {\em JHEP}, 06:128, 2014.

\bibitem{bib:HTC20}
O.~Hart, G.~{D}e Tomasi, and C.~Castelnovo.
\newblock {From compact localized states to many-body scars in the random
  quantum comb}.
\newblock {\em Phys. Rev. Research}, 2:043267, 2020.

\bibitem{bib:HVRP20}
A.~Hudomal, I.~Vasi{\'{c}}, N.~Regnault, and Z.~Papi{\'{c}}.
\newblock {Quantum scars of bosons with correlated hopping}.
\newblock {\em Communications Physics}, 3:99, 2020.

\bibitem{bib:I76}
W.~Israel.
\newblock {Thermo-field dynamics of black holes}.
\newblock {\em Phys. Lett. A}, 57:107, 1998.

\bibitem{bib:JM95}
M.~Jimbo and T.~Miwa.
\newblock {\em {Algebraic analysis of solvable lattice models}}.
\newblock American Mathematical Society, 1995.

\bibitem{bib:HMPP}
H.~Katsura, C.~Matsui, C.~Paletta, and B.~Pozsgay.
\newblock in preparation.

\bibitem{bib:K90}
T.~Kennedy.
\newblock {Exact diagonalisations of open spin-1 chains}.
\newblock {\em J. Phys. Condens. Matter}, 2:5737, 1990.

\bibitem{bib:KHN20}
V.~Khemani, M.~Hermele, and R.~Nandkishore.
\newblock {Localization from Hilbert space shattering: From theory to physical
  realizations}.
\newblock {\em Phys. Rev. B}, 101:174204, 2020.

\bibitem{bib:KBDKMZ08}
B.~Kraus, H.~P. B{\"{u}}chler, S.~Diehl, A.~Kantian, A.~Micheli, and P.~Zoller.
\newblock {Preparation of entangled states by quantum Markov processes}.
\newblock {\em Phys. Rev. A}, 78:042307, 2008.

\bibitem{bib:LMPC20}
K.~Lee, R.~Melendrez, A.~Pal, and H.~J. Changlani.
\newblock {Exact three-colored quantum scars from geometric frustration}.
\newblock {\em Phys. Rev. B}, 101:241111, 2020.

\bibitem{bib:L11}
A.~Leviatan.
\newblock {Partial Dynamical Symmetries}.
\newblock {\em Prog. Part. Nucl Phys.}, 66:93, 2011.

\bibitem{bib:LSP23}
Y.~Li, P.~Sala, and F.~Pollmann.
\newblock {Hilbert space fragmentation in open quantum systems}.
\newblock {\em Phys. Rev. Research}, 5:043239, 2023.

\bibitem{bib:LZP19}
M.~Ljubotina, M.~{\v{Z}}nidari{\v{c}}, and T.~Prosen.
\newblock {Kardar-Parisi-Zhang physics in the quantum Heisenberg magnet}.
\newblock {\em Phys. Rev. Lett.}, 122:210602, 2019.

\bibitem{bib:LDOV23}
L.~Lootens, C.~Delcamp, G.~Ortiz, and F.~Verstraete.
\newblock {Dualities in One-Dimensional Quantum Lattice Models: Symmetric
  Hamiltonians and Matrix Product Operator Intertwiners}.
\newblock {\em PRX Quantum}, 4:020357, 2023.

\bibitem{bib:M98}
Z.~Maassarani.
\newblock {The $XXC$ models}.
\newblock {\em Phys. Lett. A}, 244:160, 1998.

\bibitem{bib:M99}
Z.~Maassarani.
\newblock {Multiplicity models}.
\newblock {\em Phys. Lett. A}, 7:627, 1999.

\bibitem{bib:MN18}
J.~M. Maillet and G.~Niccoli.
\newblock {On quantum separation of variables}.
\newblock {\em J. Math. Phys.}, 59:091417, 2018.

\bibitem{bib:M03}
J.~Maldacena.
\newblock {Eternal black holes in anti-de Sitter}.
\newblock {\em JHEP}, 04:021, 2003.

\bibitem{bib:M24}
C.~Matsui.
\newblock {Exactly solvable subspaces of nonintegrable spin chains with
  boundaries and quasiparticle interactionso}.
\newblock {\em Phys. Rev. B}, 109:104307, 2024.

\bibitem{bib:MT24}
C.~Matsui and N.Tsuji.
\newblock {Exact steady states of the impurity-doped $XXZ$ spin chain coupled
  to dissipators}.
\newblock {\em J. Stat. Mech: Theor. Exp.}, 2024:033105, 2024.

\bibitem{bib:MP17}
C.~Matsui and T.~Prosen.
\newblock {Construction of the steady state density matrix and quasilocal
  charges for the spin-1/2 XXZ chain with boundary magnetic fields}.
\newblock {\em J. Phys. A: Math. Theor.}, 50:385201, 2017.

\bibitem{bib:M64}
J.~B. {McGuire}.
\newblock {Study of Exactly Soluble One‐Dimensional $N$‐Body Problems}.
\newblock {\em J. Math. Phys.}, 5:622, 1964.

\bibitem{bib:MEP16}
M.~V. Medvedyeva, F.~H.~L. Essler, and T.~Prosen.
\newblock {Exact Bethe ansatz spectrum of a tight-binding chain with dephasing
  noise}.
\newblock {\em Phys. Rev. Lett.}, 117:137202, 2016.

\bibitem{bib:MBD97}
G.~I. Menon, M.~Barma, and D.~Dhar.
\newblock {Conservation laws and integrability of a one-dimensional model of
  diffusing dimers}.
\newblock {\em J. Stat. Phys.}, 86:1237, 1997.

\bibitem{bib:MTPAS20}
A.~A. Michailidis, C.~J. T.~Z. Papi{\'{c}}, D.~A. Abanin, and M.~Serbyn.
\newblock {Stabilizing two-dimensional quantum scars by deformation and
  synchronization}.
\newblock {\em Phys. Rev. Research}, 2:022065, 2020.

\bibitem{bib:MBR22}
S.~Moudgalya, B.~A. Bernevig, and N.~Regnault.
\newblock {Quantum many-body scars and Hilbert space fragmentation: a review of
  exact results}.
\newblock {\em Rep. Prog. Phys.}, 85:086501, 2022.

\bibitem{bib:MRBR18-2}
S.~Moudgalya, S.~R. B.~A. Bernevig, and N.~Regnault.
\newblock {Entanglement of exact excited states of Affleck-Kennedy-Lieb-Tasaki
  models: Exact results, many-body scars, and violation of the strong
  eigenstate thermalization hypothesis}.
\newblock {\em Phys. Rev. B}, 98:235156, 2018.

\bibitem{bib:MRBR18}
S.~Moudgalya, S.~R. B.~A. Bernevig, and N.~Regnault.
\newblock {Exact excited states of nonintegrable models}.
\newblock {\em Phys. Rev. B}, 98:235155, 2018.

\bibitem{bib:MM22}
S.~Moudgalya and O.~I. Motrunich.
\newblock {Hilbert Space Fragmentation and Commutant Algebras}.
\newblock {\em Phys. Rev. X}, 12:011050, 2022.

\bibitem{bib:MBBFR20}
S.~Moudgalya, E.~O'Brien, B.~A. Bernevig, P.~Fendley, and N.~Regnault.
\newblock {Large classes of quantum scarred Hamiltonians from matrix product
  states}.
\newblock {\em Phys. Rev. B}, 102:085120, 2020.

\bibitem{bib:MPNRB21}
S.~Moudgalya, A.~Prem, R.~Nandkishore, N.~Regnault, and B.~A. Bernevig.
\newblock {Chapter 7: Thermalization and Its Absence within Krylov Subspaces of
  a Constrained Hamiltonian}.
\newblock In {\em Memorial Volume for Shoucheng Zhang}, page 147, 2021.

\bibitem{bib:MRB20}
S.~Moudgalya, N.~Regnault, and B.~A. Bernevig.
\newblock {$\eta$-pairing in Hubbard models: From spectrum generating algebras
  to quantum many-body scars}.
\newblock {\em Phys. Rev. B}, 102:085140, 2020.

\bibitem{bib:N04}
R.~I. Nepomechie.
\newblock {Bethe ansatz solution of the open XXZ chain with nondiagonal
  boundary terms}.
\newblock {\em J. Phys. A: Math. Gen.}, 37:433, 2004.

\bibitem{bib:DBCK20-2}
N.~O’Dea, F.~Burnell, A.~Chandran, and V.~Khemani.
\newblock {From tunnels to towers: Quantum scars from Lie algebras and
  $q$-deformed Lie algebras}.
\newblock {\em Phys. Rev. Research}, 2:043305, 2020.

\bibitem{bib:DBCK20}
N.~O’Dea, F.~Burnell, A.~Chandran, and V.~Khemani.
\newblock {Quasisymmetry Groups and Many-Body Scar Dynamics}.
\newblock {\em Phys. Rev. Research}, 2:043305, 2020.

\bibitem{bib:PPN19}
S.~Pai, M.~Pretko, and R.~M. Nandkishore.
\newblock {Localization in Fractonic Random Circuits}.
\newblock {\em Phys. Rev. X}, 9:021003, 2019.

\bibitem{bib:PPPK20}
K.~Pakrouski, P.~N. Pallegar, F.~K. Popov, and I.~R. Klebanov.
\newblock {Many-Body Scars as a Group Invariant Sector of Hilbert Space}.
\newblock {\em Phys. Rev. Lett.}, 125:230602, 2020.

\bibitem{bib:PPPK21}
K.~Pakrouski, P.~N. Pallegar, F.~K. Popov, and I.~R. Klebanov.
\newblock {Group theoretic approach to many-body scar states in fermionic
  lattice models}.
\newblock {\em Phys. Rev. Research}, 3:043156, 2021.

\bibitem{bib:PP24}
C.~Paletta and T.~Prosen.
\newblock {Integrability of open boundary driven quantum circuits}.
\newblock arXiv:2406.12695 [quant-ph], 2024.

\bibitem{bib:P18}
B.~Pozsgay.
\newblock {Overlaps with arbitrary two-site states in the XXZ spin chain}.
\newblock {\em J. Stat. Mech.}, 2018:053103, 2018.

\bibitem{bib:P08}
T.~Prosen.
\newblock {Third quantization: a general method to solve master equations for
  quadratic open Fermi systems}.
\newblock {\em New J. Phys.}, 10:043026, 2008.

\bibitem{bib:P11}
T.~Prosen.
\newblock {Open $XXZ$ Spin Chain: Nonequilibrium Steady State and a Strict
  Bound on Ballistic Transport}.
\newblock {\em Phys. Rev. Lett}, 106:217206, 2011.

\bibitem{bib:RLF21}
J.~Ren, C.~Liang, and C.~Fang.
\newblock {Quasisymmetry Groups and Many-Body Scar Dynamics}.
\newblock {\em Phys. Rev. Lett.}, 126:120604, 2021.

\bibitem{bib:RLF22}
J.~Ren, C.~Liang, and C.~Fang.
\newblock {Deformed symmetry structures and quantum many-body scar subspaces}.
\newblock {\em Phys. Rev. Research}, 4:013155, 2022.

\bibitem{bib:SRVKP20}
P.~Sala, T.~Rakovszky, R.~Verresen, M.~Knap, and F.~Pollmann.
\newblock {Ergodicity Breaking Arising from Hilbert Space Fragmentation in
  Dipole-Conserving Hamiltonians}.
\newblock {\em Phys. Rev. X}, 10:011047, 2020.

\bibitem{bib:Sato95}
R.~Sato.
\newblock {Bethe ansatz calculations for quantum spin chains with partial
  integrability}.
\newblock {\em J. Phys. Soc. Jpn.}, 64:2837, 1995.

\bibitem{bib:SI19}
M.~Schecter and T.~Iadecola.
\newblock {Weak Ergodicity Breaking and Quantum Many-Body Scars in Spin-$1$
  $XY$ Magnets}.
\newblock {\em Phys. Rev. Lett.}, 123:147201, 2019.

\bibitem{bib:SK19}
N.~Shibata and H.~Katsura.
\newblock {Dissipative spin chain as a non-Hermitian Kitaev ladder}.
\newblock {\em Phys. Rev. B}, 99:174303, 2019.

\bibitem{bib:SYK20}
N.~Shibata, N.~Yoshioka, and H.~Katsura.
\newblock {Onsager’s Scars in Disordered Spin Chains}.
\newblock {\em Phys. Rev. Lett.}, 124:180604, 2020.

\bibitem{bib:S85}
E.~K. Sklyanin.
\newblock {The quantum Toda chain}.
\newblock In {\em {Non-Linear Equations in Classical and Quantum Field
  Theory}}, page 196, 1985.

\bibitem{bib:S88}
E.~K. Sklyanin.
\newblock {Boundary conditions for integrable quantum systems}.
\newblock {\em J. Phys. A: Math. Gen.}, 21:2375, 1988.

\bibitem{bib:S90}
E.~K. Sklyanin.
\newblock {FUNCTIONAL BETHE ANSATZ}.
\newblock In {\em {Integrable and Superintegrable Systems}}, page~8, 1990.

\bibitem{bib:S92}
E.~K. Sklyanin.
\newblock {Quantum inverse scattering method. Selected topics}.
\newblock In {\em {Quantum Group and Quantum Integrable Systems}}, page~63,
  1992.

\bibitem{bib:S92-2}
E.~K. Sklyanin.
\newblock {Separation of variables in the classical integrable $SL(3)$ magnetic
  chain}.
\newblock {\em Commun. Math. Phys.}, 150:181, 1992.

\bibitem{bib:S95}
E.~K. Sklyanin.
\newblock {Separation of Variables: New Trends}.
\newblock {\em Prog. Theor. Phys. Suppl.}, 118:35, 1995.

\bibitem{bib:S96}
E.~K. Sklyanin.
\newblock {Separation of variables in the quantum integrable models related to
  the Yangian $y[sl(3)]$}.
\newblock {\em J. Math. Sci.}, 80:1861, 1996.

\bibitem{bib:STF79}
E.~K. Sklyanin, L.~A. Takhtadzhyan, and L.~D. Faddeev.
\newblock {Quantum inverse problem method. I}.
\newblock {\em Theor. Math. Phys.}, 40:194, 1979.

\bibitem{bib:SP98}
A.~I. Solomon and K.~A. Penson.
\newblock {Coherent pairing states for the Hubbard model}.
\newblock {\em J. Phys. A: Math. Gen.}, 31:L355, 1998.

\bibitem{bib:TMBJ20}
J.~Tindall, C.~S.~M. {n}oz, B.~Bu\v{c}a, and D.~Jaksch.
\newblock {Quantum synchronisation enabled by dynamical symmetries and
  dissipation}.
\newblock {\em New J. Phys.}, 22:013026, 2020.

\bibitem{bib:TMASP18-2}
C.~J. Turner, A.~A. Michailidis, D.~A. Abanin, M.~Serbyn, and Z.~Papi\'{c}.
\newblock {Quantum scarred eigenstates in a Rydberg atom chain: Entanglement,
  breakdown of thermalization, and stability to perturbations}.
\newblock {\em Phys. Rev. B}, 98:155134, 2018.

\bibitem{bib:TMASP18}
C.~J. Turner, A.~A. Michailidis, D.~A. Abanin, M.~Serbyn, and Z.~Papi\'{c}.
\newblock {Weak ergodicity breaking from quantum many-body scars}.
\newblock {\em Nature Phys.}, 14:745, 2018.

\bibitem{bib:V20}
E.~Vernier.
\newblock Mixing times and cutoffs in open quadratic fermionic systems.
\newblock {\em SciPost Phys.}, 20:049, 2010.

\bibitem{bib:WM24}
C.~W. W{\"{a}}chtler and J.~E. Moore.
\newblock {Topological Quantum Synchronization of Fractionalized Spins}.
\newblock {\em Phys. Rev. Lett.}, 132:196601, 2024.

\bibitem{bib:WYCS15}
Y.~Wang, W.-L. Yang, J.~Cao, and K.~Shi.
\newblock {\em {Off-Diagonal Bethe Ansatz for Exactly Solvable Models}}.
\newblock Springer, 2015.

\bibitem{bib:Y67}
C.~N. Yang.
\newblock {Some Exact Results for the Many-Body Problem in one Dimension with
  Repulsive Delta-Function Interaction}.
\newblock {\em Phys. Rev. Lett.}, 19:1312, 1967.

\bibitem{bib:ZF21}
L.~Zadnik and M.~Fagotti.
\newblock {The Folded Spin-1/2 XXZ Model: I. Diagonalisation, Jamming, and
  Ground State Properties}.
\newblock {\em SciPost Phys. Core}, 4:10, 2021.

\bibitem{bib:ZKMS97}
S.~Zhang, M.~Karbach, G.~M{\"{u}}ller, and J.~Stolze.
\newblock {Charge and spin dynamics in the one-dimensional $t$-$J_z$ and
  $t$-$J$ models}.
\newblock {\em PRX Quantum}, 55:6491, 1997.

\bibitem{bib:ZCYSW14}
X.~Zhang, J.~Cao, W.-L. Yang, K.~Shi, and Y.~Wang.
\newblock {Exact solution of the one-dimensional super-symmetric $t$–$J$
  model with unparallel boundary fields}.
\newblock {\em J. Stat. Mech.}, 2014:P04031, 2014.

\bibitem{bib:ZSMK21}
H.~Zhao, A.~Smith, F.~Mintert, and J.~Knolle.
\newblock {Orthogonal Quantum Many-Body Scars}.
\newblock {\em Phys. Rev. Lett.}, 127:150601, 2021.

\bibitem{bib:ZE20}
A.~A. Ziolkowska and F.~H.~L. Essler.
\newblock {Yang-Baxter integrable Lindblad equations}.
\newblock {\em SciPost Phys.}, 8:044, 2020.

\end{thebibliography}

\end{document}